\documentclass{aa}
\usepackage{graphicx}
\usepackage{txfonts}
\usepackage{natbib}

\begin{document}
\title{Radio observational constraints on Galactic 3D-emission models}
\author{X. H.~Sun\inst{1}\fnmsep\thanks{Current address: National Astronomical 
           Observatories, Chinese Academy of Sciences, Jia-20 Datun Road, 
           Chaoyang District, Beijing 100012, China. E-mail: xhsun@bao.ac.cn}, 
        W.~Reich\inst{1}, 
        A.~Waelkens\inst{2} 
        \and T. A.~En{\ss}lin\inst{2}}

\offprints{Wolfgang Reich\\ \email{wreich@mpifr-bonn.mpg.de}}

\institute{Max-Planck-Institut f\"{u}r Radioastronomie, Auf dem H\"{u}gel 69, 
           53121 Bonn, Germany 
      \and Max-Planck-Institut f\"{u}r Astrophysik, Karl-Schwarzschild-Str. 1, 
           85741 Garching, Germany}

\date{Received ; accepted }

\abstract 
{Our position inside the Galaxy requires 3D-modelling to obtain the 
distribution of the Galactic magnetic field, cosmic-ray (CR) electrons and 
thermal electrons. 
}
{Our intention is to find a Galactic 3D-model which agrees best with
available radio observations.}
{We constrain simulated all-sky maps in total intensity, linear
 polarization, and rotation measure (RM) by observations. For the
 simulated maps as a function of frequency we integrate in $15\arcmin$
 wide cones the emission along the line of sight calculated from
 Galactic 3D-models. We test a number of large-scale magnetic field
 configurations and take the properties of the warm interstellar
 medium into account.}
{From a comparison of simulated and observed maps we are able to
constrain the regular large-scale Galactic magnetic field in the disk
and the halo of the Galaxy. The local regular field is 2~$\mu$G and
the average random field is about 3~$\mu$G. The known local excess of
synchrotron emission originating either from enhanced CR electrons or
random magnetic fields is able to explain the observed high-latitude
synchrotron emission. The thermal electron model (NE2001) in
conjunction with a proper filling factor accounts for the observed
optically thin thermal emission and low frequency absorption by
optically thick emission.  A coupling factor between thermal electrons
and the random magnetic field component is proposed, which in addition
to the small filling factor of thermal electrons increases small-scale
RM fluctuations and thus accounts for the observed depolarization at
1.4~GHz.}
{We conclude that an axisymmetric magnetic disk field configuration
with reversals inside the solar circle fits available observations
best. Out of the plane a strong toroidal magnetic field with different
signs above and below the plane is needed to account for the observed
high-latitude RMs. The large field strength is a consequence of the
small thermal electron scale height of 1~kpc, which also limits the CR
electron extent up to a height of 1~kpc not to contradict with the
observed synchrotron emission out of the plane. Our preferred 3D-model 
fits the observed Galactic total intensity and polarized emission better 
than other models over a wide frequency range and also agrees with 
the observed RM from extragalactic sources.}

\keywords {polarization -- Radiation mechanisms: non-thermal -- 
Radiation mechanisms: thermal -- ISM: magnetic field -- ISM: structure -- 
Galaxy: structure}

\titlerunning{Galactic 3D-models}
\authorrunning{X.H.~Sun et al.}

\maketitle

\section{Introduction}

Synchrotron radiation is the major component of the Galactic radio
continuum emission from centimetre to longer wavelengths. The diffuse
Galactic synchrotron emission originates from relativistic electrons
gyrating in the magnetic field. An additional smaller contribution is
observed from supernova remnants (SNRs) and pulsar wind nebulae
located in the Galactic plane, where cosmic ray (CR) electrons are
accelerated to high energies. Synchrotron emission is intrinsically
highly polarized, but Faraday rotation effects in the magnetized
interstellar medium along the line of sight and also beam averaging
effects reduce the observed percentage polarization to much smaller
values. It is crucial to understand the regularity, strength and
spatial structure of the magnetic field in our Galaxy, which are still
under discussion although the first detection of Galactic polarization
was several decades ago \citep{wsp62,wsb+62}. Our position inside the
Galactic disk at about 8.5~kpc distance from the Galactic centre
requires modelling of the magnetic field and CR electrons to interpret
observed synchrotron maps. Polarization maps are influenced by Faraday
rotation effects along the line of sight and thus require to take the
distribution of thermal electrons in the Galaxy into account as
well. Important information on the Galactic magnetic field strength
and direction comes from new rotation measure (RM) data. All these
information must be commonly considered to constrain Galactic
3D-models in order to find the most realistic one. This paper presents
a general attempt in that direction.

Previous 3D-models for the magnetic field, CR electrons and thermal
electrons were derived individually from selected data sets. Their
general relevance can be established by combining these models
to explain other, in particular new observational data, for instance
from WMAP.  Here we consider RMs from extragalactic sources (EGSs),
thermal and non-thermal intensities and linearly polarized emission
together, which proves to give stringent constraints to combined
3D-models. We note that simulated maps are also required as a robust
reference for the technical design of future telescope projects like
the SKA or as inputs for data reduction pipelines like that for the
PLANCK mission.

The paper is organized as follows: we review the status of Galactic
models in Sect. 2, describe the method of our simulations in Sect. 3,
summarize available observations used to constrain the 3D-models in
Sect. 4 and briefly describe numerical details of the simulations in
Sect. 5. The 3D-model results are presented in Sect. 6. Some general
problems relevant to the modelling are discussed in
Sect. 7. Conclusions are presented in Sect. 8.

\section{Status of Galactic 3D-models}

Many attempts have been made to recover the structure of the Galactic
synchrotron emission. \citet{bkb85} deconvolved the 408~MHz all-sky
total intensity map \citep{hssw82} and modelled the emission as a
thick disk with a kpc scale height contributing about 90\% of the
total emission plus an embedded thin disk with a scale height of about
200~pc, which is consistent with the results by
\citet{pko+81a,pko+81b} using the same data. They adapt a spiral
pattern for the emissivity and assume comparable regular and random
magnetic field strength.  \citet{ft95} present observational evidence
for an excess of synchrotron emissivity within a few hundred parsec of
the solar system based on low-frequency absorption data towards a
number of large HII regions. A recent study of local Faraday screens
by \citet{wr04} seems to confirm this result. Since local emission is
also visible at high Galactic latitudes the thick disk assumption
might be re-discussed.

The observed spectrum of Galactic synchrotron emission shows complex
spatial variations due to processes such as energy losses, diffusion
of CR electrons or variations of the magnetic field strength, which
makes a modelling of the spectral behaviour rather difficult. For the
northern sky \citet{rr88a,rr88b} found spectral variations
($T_\nu\propto\nu^\beta$, with $T_\nu$ the brightness temperature at
frequency $\nu$ and $\beta$ the temperature spectral index) in the
range $\beta \sim -3.1$ to $\beta \sim -2.5$ between 408~MHz and
1420~MHz.  At lower frequencies the spectrum is
flatter. \citet{rcls99} find spectral variations between $\beta \sim
-2.55$ and $\beta \sim -2.4$ between 22~MHz and 408~MHz. At higher
frequencies the synchrotron spectrum steepens, but the amount is
unclear so far. For the 1420~MHz and 22.8~GHz surveys \citet{rrt04}
calculate steeper spectra towards higher latitudes than below
1420~MHz. However, the 22.8~GHz (K-band) WMAP data \citep{hnb+07} may
contain an unknown fraction of anomalous dust emission \citep{ddb+06}
and thus the synchrotron spectrum might be steeper than the total
emission spectrum. For any modelling it implies that the assumption of
a constant spectral index is a simplification.

Synchrotron emission is generated by CR electrons in the Galactic
magnetic field, which are considered as independent. We are
unfortunately located in the disk of our Galaxy, which makes the
observation of the large-scale field difficult. From observations of
other spiral galaxies it is known that the field basically follows a
spiral pattern \citep{bbm+96}. There are many methods to obtain
information about the Galactic magnetic field observationally (see
e.g. \citet{hw02}). A powerful method is the analysis of pulsar RMs,
which indicates from recent data that the field direction is opposite
in the arm and the inter-arm regions \citep{hml+06}. However, RM data
from EGSs can be interpreted with less reversals
\citep{btwm03,hgb+06,bhg+07}. The local regular field strength is
about 2~$\mu$G and increases to about 4~$\mu$G at about 3~kpc from the
Galactic centre according to \citet{hml+06}, where no coupling between
the magnetic field strength and the thermal electron density is
assumed.  \citet{smr00} estimated the total local field to be about
6~$\mu$G. The Galactic random field component was estimated to 
be about 6~$\mu$G \citep{hml+06}. This is about a factor of 2 to 3
larger than the regular field component. This ratio is larger than it
was assumed in earlier models.

The distribution of Galactic CR electrons has still to be
established. The emission at radio frequencies originates from GeV
electrons, where a power law distribution is a good
approximation. SNRs are the acceleration sites and thus the sources of
CR electrons. However, the propagation of CRs in our Galaxy is very
complex and modelling efforts like GALPROP \citep{sm98} are required.

The properties for the magnetized thermal interstellar medium are
crucial for the interpretation of polarization observations, because
Faraday effects strongly modulate the polarization emission along the
line of sight and make modelling of Galactic polarization quite
challenging. At low frequencies Faraday effects might be strong enough
to entirely smear out distant polarization \citep{sbs+98}. For the
Galactic large-scale thermal component the NE2001 model based on
pulsar dispersion measures \citep{cl02} represents the best
knowledge. However, additional Faraday effects come from HII regions
and also from Faraday screens hosting strong regular fields, but a low
thermal electron density \citep{wr04,shr+07,rei07}, and might
influence the observations by a yet unknown amount.

\begin{figure*}[!htbp]
\centering
\includegraphics[width=13cm,angle=-90]{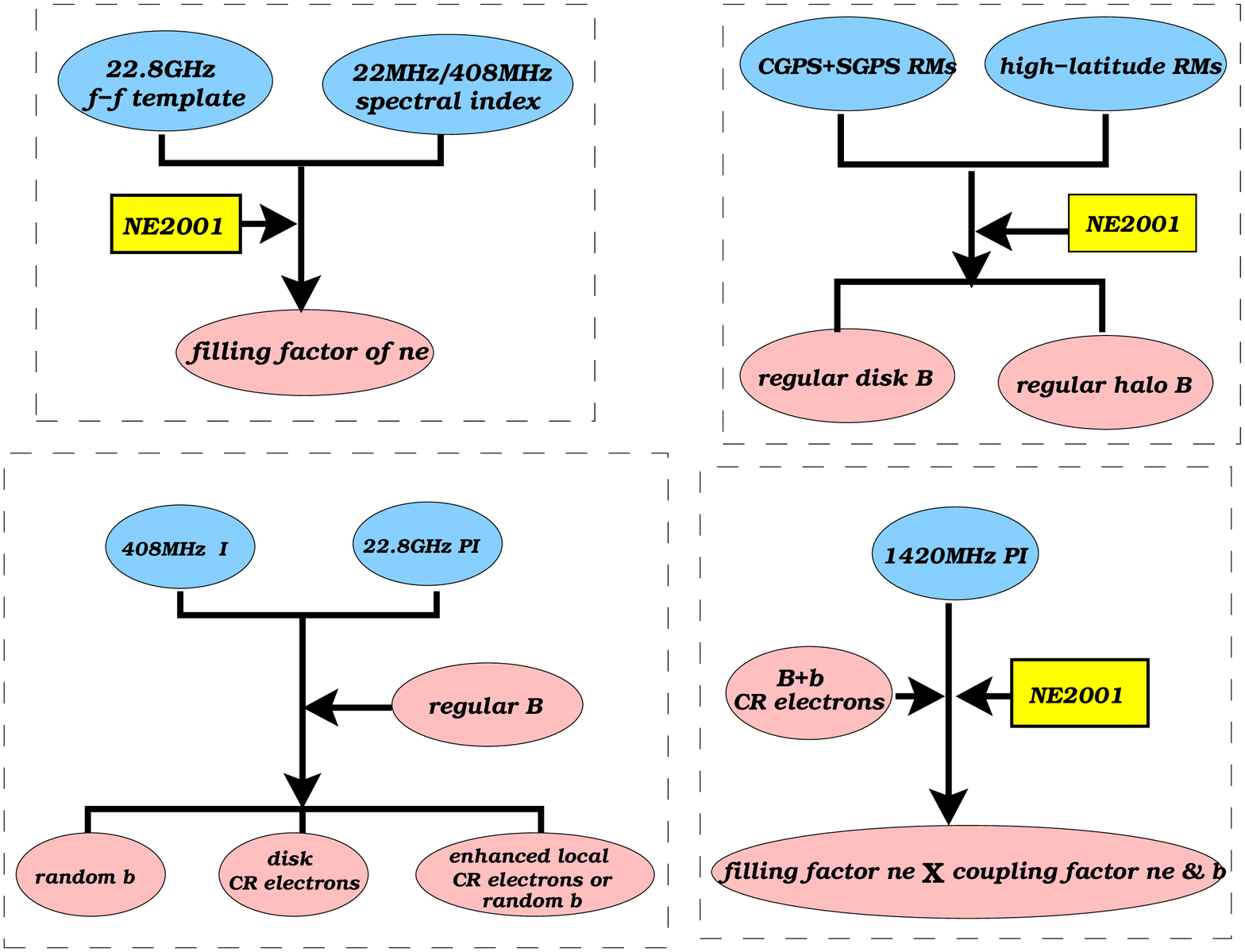}
\caption{Overview on the different steps applied to find the Galactic
3D-emission components and their properties. Here f-f stands for
free-free emission, ne for thermal electons, $\rm \bf{B}~(\bf{b})$
vectors for the regular (random) magnetic field component, and I (PI)
for total (polarized) intensity. CGPS and SGPS mean the Canadian and
the Southern Galactic Plane Survey, respectively. The blue color is 
for observational data, the pink for the model and the yellow for 
NE2001. }
\label{simprocess}
\end{figure*}

\section{The method applied}

The observables are RM data, total and polarized intensity maps at
different frequencies. Polarization angles and percentage polarization
can be used also for constraints. The goal is to obtain the
thermal electron, CR electron and magnetic field 3D-distribution in
the Galaxy which agrees best with the observations.

RMs depend on the regular magnetic field component and the thermal
electron density along the line of sight. The perpendicular magnetic
field component and the CR electron density determine the total
intensity synchrotron emission. Observed polarized intensity depends
on the regular fraction of the magnetic field, but gets modified as a
function of frequency by the line of sight magnetic field components
and the thermal electron density. To achieve a 'best model' an
optimized quantity like $\chi^2$ should be used to adjust the model
parameter. This is beyond the scope of the present attempt and also
clearly needs more observations than actually available for a well
established model.

We tackle the 3D-modelling problem by finding the model parameters in
four steps as shown in Fig.~\ref{simprocess}. The modelling is based
on the thermal electron density distribution as it is described by the
widely accepted NE2001 model \citep{cl02}. After considering a proper
filling factor for the thermal electron distribution of the NE2001
model we are able to reproduce the observed optically thin free-free
emission and the effect of absorption from the optically thick
gas. This indicates that the NE2001 model is valid in general,
although a larger scale height is favoured by our modelling results as
discussed later. From RM data of EGSs we are able to obtain the
regular magnetic field direction and its strength separately for the
Galactic plane and towards higher latitudes by keeping the thermal
electron model parameters from NE2001 fixed. From the observed 408~MHz
total intensity and the 22.8~GHz polarized intensity we constrain the
random magnetic field component and the large-scale distribution of CR
electrons using the regular magnetic field component obtained in the
step before.  If the random field component is increased or decreased
the CR electron density must be reduced or enlarged accordingly to fit
the total intensity emission. Both components must be matched to fit
the observed polarized intensity.

The local excess of synchrotron emission is considered in our model,
which is able to account for the observed high-latitude emission as
well. At that point our model is able to represent the total intensity
synchrotron emission and the intrinsic polarization properties as
observed at high frequencies and is in general agreement with the
observed extragalactic RM distribution. In the last step, the observed
depolarization properties at 1.4~GHz are accounted for by introducing
a coupling factor between the thermal electron density and the
magnetic field strength on small-scales in addition to the thermal
filling factor.

\section{Available observations constraining the 3D-models}

To constrain the simulated maps from Galactic 3D-models we consider
various observed all-sky data sets. These maps contain the integrated
line of sight information of large-scale Galactic emission, but also
show all kinds of local foreground structures and distinct source
complexes, which are not the subject of modelling in our case. We note
that it is not always clear how to distinguish and to correctly
separate the observed structures into local and global emission
components. Unresolved strong sources can be easily identified and are
not represented by the models. Large-scale local features like the
giant radio loops with diameters up to $100\degr$ clearly show up as
distinct feature in the total intensity and the polarization maps and
are also not included in the model, although they are difficult to
separate from the observed maps. The large-scale highly polarized
``Fan region'' is difficult to trace in total intensities, but
certainly contributes to the observed emission. Distinct SNRs and
large HII region complexes are concentrated in the Galactic plane and
are excluded from the modelling, but their position is known from
high-resolution surveys along the Galactic plane. Faraday screens of
considerable size exist in the interstellar medium, which are weak
emitters, but reflect their existence by their large RMs by rotating
the polarization angles of the background emission. Only a very
densely sampled RM grid of sources or multi-frequency observations of
diffuse Galactic polarization can reveal such features. These data are
not available, but their importance is widely accepted. ``Key science
projects'' for LOFAR and the future SKA were defined \citep{gbf+04},
which exactly have these aims.  Faraday screens might influence our
modelling results to an unknown extent. All the uncertainties in the
recognition of local or distinct features limits any Galactic
modelling attempt to the reconstruction of the large-scale Galactic
structures.

\begin{figure}[!htbp]
\centering
\includegraphics[width=9cm]{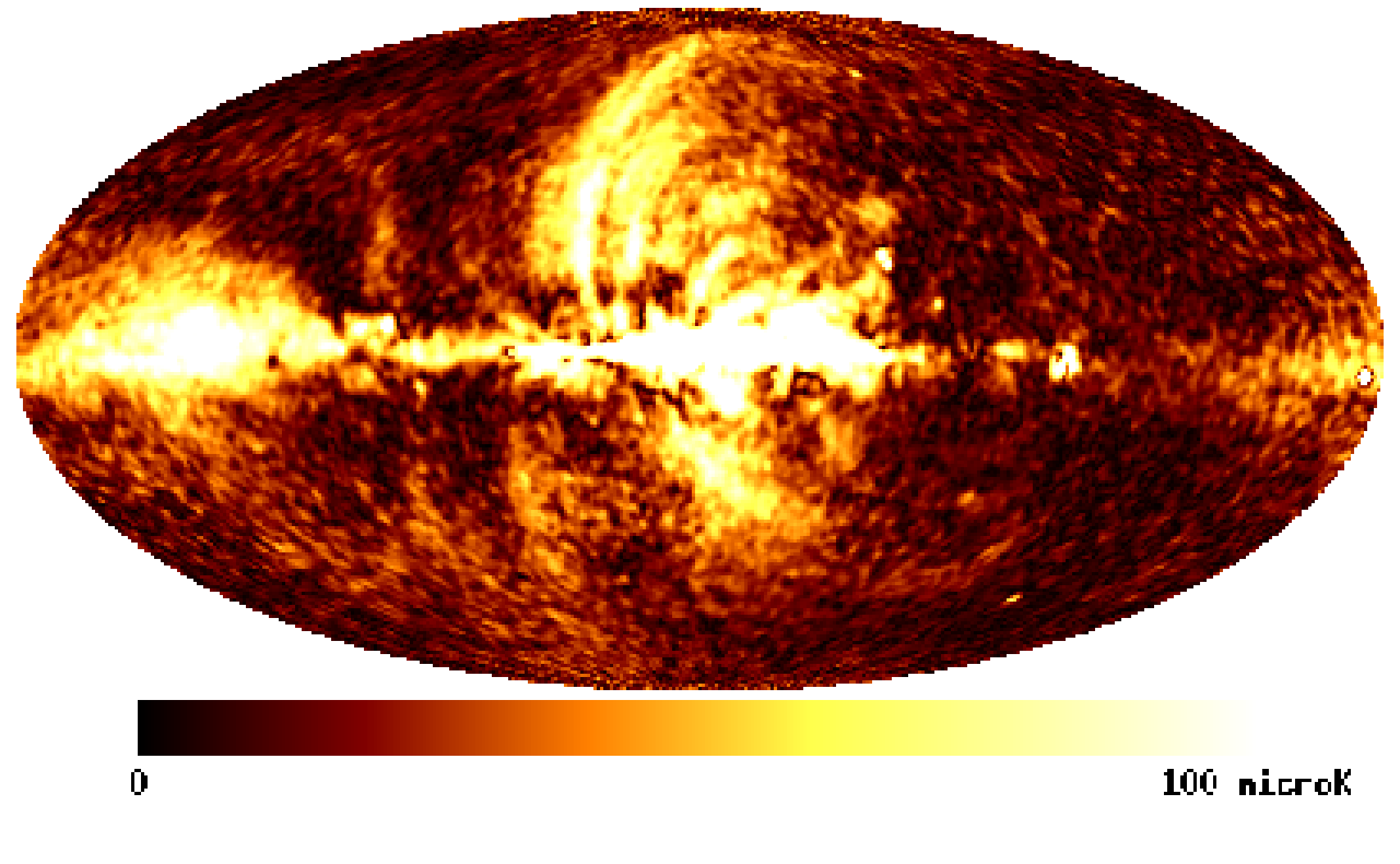}
\includegraphics[width=9cm]{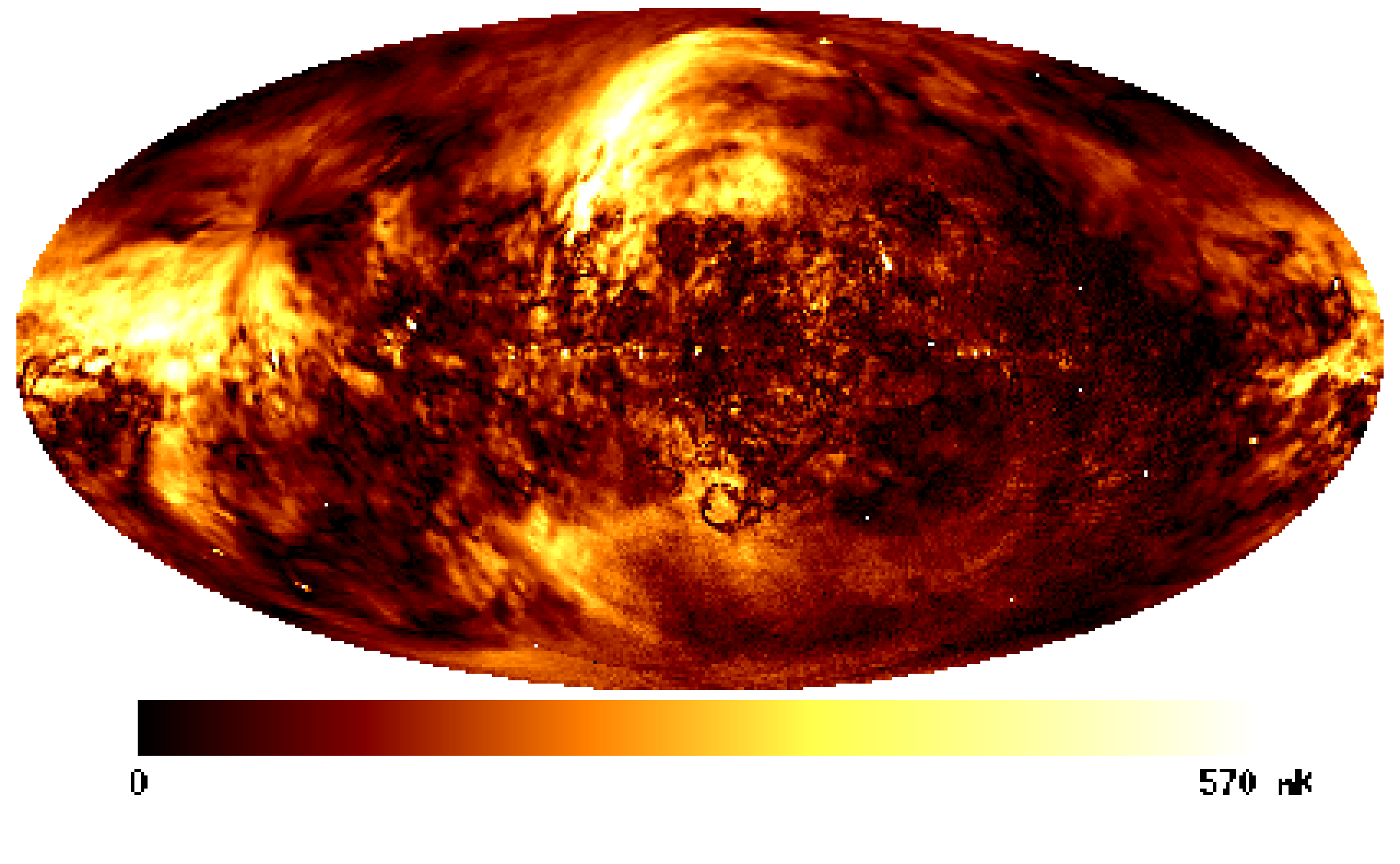}
\caption{The WMAP 22.8~GHz all-sky polarized intensity map (upper
panel) and the 1.4~GHz all-sky polarized intensity map (lower panel).
The polarized intensities are shown greyscale coded from 0 to
100~$\mu$K for 22.8~GHz and from 0 to 570~mK for 1.4~GHz.}
\label{pik}
\end{figure}

\subsection{All-sky total intensity maps}

All-sky total intensity surveys up to a few GHz, which are adjusted to
an absolute zero-level, are essential to obtain the required
large-scale knowledge of the Galactic synchrotron emission. Carrying
out such surveys is rather time-consuming and therefore these surveys
have relatively low angular resolution. Available all-sky surveys have
been reviewed by \citet{rei03}. Among these the 408~MHz survey
\citep{hssw82} and the 1420~MHz survey \citep{rei82,rr86,rtr01} have
the highest sensitivity and an angular resolution down to $0\fdg6$ and
thus are well suited for our purpose. 

Galactic radio emission at frequencies below a few GHz is dominated by
synchrotron emission, in particularly out of the plane, relative to
free-free emission. Thermal dust emission and probably spinning dust
emission are becoming important for frequencies higher than 5~GHz or
10~GHz. The 408~MHz total intensity map is often used as a
``synchrotron template'' and in fact significant thermal emission is
confined to the Galactic plane. At 1420~MHz the thermal fraction in
the plane increases to about 40\% in the inner Galaxy
\citep{rr88b}. \citet{pddg05} estimated a thermal contribution with a
maximum of about 68\% for the region of $20\degr<l<30\degr$ and
$|b|<1\fdg5$, where discrete HII regions contribute about 9\%. Note
that these estimates are based on the separation of thermal and
non-thermal components assuming appropriate spectral indices. However,
the spectral index of synchrotron emission is uncertain and might vary
with direction, although these estimates are largely consistent with
the method to separate both components based on a relation between
infrared and thermal emission \citep{bho89}. It is certainly worth to
refine the separation of thermal and non-thermal emission, but this is
beyond the scope of the present paper.

\citet{ddd03} have constructed an extinction corrected free-free
emission template for about 95\% of the sky based on H$\alpha$
surveys of the northern sky \citep{hrt+03} and of the southern sky
\citep{gmrv01} by using dust emission data \citep{sfd98} to calculate
the extinction correction. \citet{hnb+07} used the H$\alpha$ template
by \citet{fin03} as a prior estimate in the maximum entropy method
analysis of the 3-year WMAP data to construct the free-free emission
template. This template might contain some leakage from other
components such as synchrotron emission and anomalous dust emission. 
However, actually it probably represents the best estimate for low
latitude emission. We will refer to the free-free emission map at
22.8~GHz from \citet{hnb+07} for comparison with the modelled thermal
emission. The \citet{ddd03} template cannot be used for the Galactic
plane area because of the large and uncertain extinction there.

All radio maps show a clear concentration of emission along the
Galactic plane. As mentioned above, local structures out of the plane
are rather dominant features. The most striking protrusion emerging at
$l \sim 30\degr$ out of the plane is called the North Polar Spur
(NPS), which can be traced up to very high latitudes. This spur is a
part of the giant Loop I, which is believed to be a local old
supernova remnant at a distance of about 200~pc \citep{ber73}.  A
recent new model by \citet{wol07} tries to explain the NPS in total
intensity and linear polarization by local synchrotron emitting shells
with a swept-up magnetic field. A number of weaker loops of similar
size exist as discussed by \citet{bhs71}.

\subsection{All-sky polarization maps}

The status of Galactic polarization surveys has been recently reviewed
by \citet{rei07}. For the interpretation of polarization surveys the
correct absolute zero-level is of particular importance as it was
demonstrated by \citet{rfr+04}. The only ground based all-sky
polarization map exists at 1.4~GHz (Reich et al., in prep.) combining
the northern sky survey carried out with the DRAO 26-m telescope
\citep{wlrw06}, which is tied to the absolute level of the 1.4~GHz
Leiden survey \citep{bs76}, and the Villa Elisa southern sky survey
\citep[][Testori et al., in prep.]{trr04}, which has a large overlap
with the DRAO survey. At this frequency the depolarization effects are
already very strong at latitudes below about $30\degr$, especially
towards the Galactic plane as can be seen from Fig.~\ref{pik}. Thus
the distance up to which polarized emission can be observed at 1.4~GHz
with an angular resolution of $36\arcmin$ is small compared to the
line of sight through the Galaxy. On the other hand the WMAP 22.8~GHz
polarized intensity map \citep{phk+07} is almost unaffected by Faraday
depolarization effects and therefore should represent the intrinsic
synchrotron properties of the Galaxy, since dust polarization is
believed to be entirely unimportant at that frequency \citep{phk+07}.

As it can be seen from the 22.8~GHz map there are strong imprints of
Loop I, which shows strong polarization along the NPS and also below
the Galactic plane. Loop I is very local as mentioned in the previous
section and therefore can be also traced for large sections at
1.4~GHz. Another very bright polarization feature called the ``Fan
region'' is centered at Galactic longitude of about 140$\degr$ and was
recognized as a local feature at a distance of about 500~pc
\citep{ws74,bs76}. However, a new analysis based on the DRAO 1.4~GHz
survey suggests that it might extend from about 500~pc to about 2~kpc
\citep{wlrw06}. We take both structures as distinct local features, as
they are very bright at 1.4~GHz contrary to the polarized emission
from the Galactic plane. The modelling of these distinct features is
beyond the scope of our simulation. They will not influence our
modelling of the large-scale Galactic structures.
   
Although external Faraday depolarization is very small at 22.8~GHz we
find a low percentage polarization at 22.8~GHz in general. This is in
agreement with the results of \citet{kog+07}. We divided the polarized
intensity map by the synchrotron emission template provided by the
WMAP team.  The polarization percentage in the Galactic plane is
typically below 2\%. At higher latitudes typical values between 12\%
and 30\% are derived, which are rather similar to those found at
1.4~GHz. Of course, the 22.8~GHz percentage polarizations may be
systematically underestimated in case of the existence of a
significant spinning dust component in the total intensity
map. Despite of this uncertainty the intrinsic synchrotron
polarization will stay significantly below the maximum polarization
value for a perfect regular field. This might indicate that the random
magnetic field exceeds the regular magnetic field component. Otherwise
the superposition of a number of regular magnetic field components
along the line of sight with different orientations may have the same
effect.
  
Comparing the 22.8~GHz and the 1.4~GHz polarized emission we see
strong depolarization effects along the Galactic plane except for the
``Fan region''. This depolarization is attributed to Faraday effects
from the warm ionized medium. However, it is not immediately clear,
whether the diffuse medium or sources like HII regions or Faraday
screens play the main role in this process. We will show below that
the depolarization calculated from the diffuse thermal component as
provided by the NE2001 model is insufficient to explain the 1.4~GHz
polarization observations.

The polarized intensity can be used to constrain the strength of the
regular magnetic field. As can be seen from the 22.8~GHz polarization
map (Fig.~\ref{pik}) the main characteristics are (1) strong
polarization towards the Galactic centre; (2) the distribution of
polarized emission along Galactic longitude is asymmetric with respect
to the Galactic centre. Depolarization can be neglected at 22.8~GHz
and therefore the polarization angles could also be used to check the
orientation of the magnetic field. However, our simulations show that
any spiral magnetic field configuration gives rather similar
polarization angle patterns. Therefore we do not use polarization
angles to decide on the field configuration as it was previously done
by \citet{phk+07}. As shown below this model is inconsistent with
other observations.

\begin{figure}[!htbp]
\centering
\includegraphics[width=9cm]{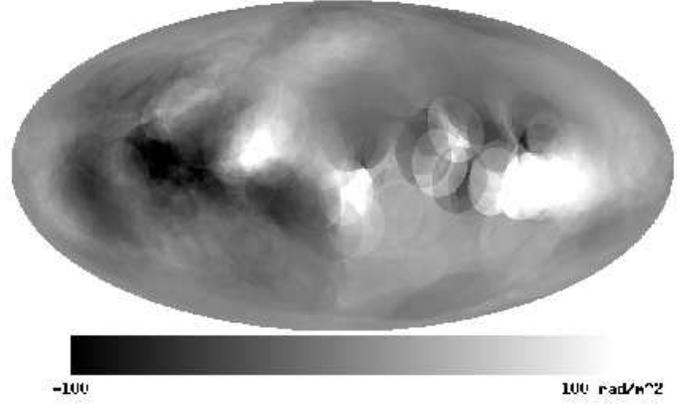}
\caption{Synthesized RM map constructed from published and new RM data
of EGSs (Han et al., in prep.). The preliminary new RM data set from
the Effelsberg survey was provided by JinLin Han. The RM map does not
include Galactic plane RM data from the Canadian (CGPS) or the
Southern Galactic Plane (SGPS) surveys.}
\label{effrm}
\end{figure}

\subsection{Rotation measures of extragalactic sources}

When a linearly polarized electromagnetic wave passes a magnetized
thermal medium its polarization angle experiences rotation, called
Faraday rotation. The rotation is proportional to the square of the
wavelength and the coefficient is the RM, which can be written as
\begin{equation}
RM=0.81\int_0^D n_e(l) B_\parallel(l) {\rm d}l
\end{equation}    
where the integral range is along the line of sight from the observer
to the source at a distance $D$ measured in pc. $n_e$ per cm$^{-3}$ is
the thermal electron density and $B_\parallel$ in $\mu G$ is the
magnetic field component along the line of sight in the intervening
medium.  RM is positive when the magnetic field points towards us.

RMs of EGSs reflect the integrated effect of the magnetized medium
throughout the Galaxy and can be used for a diagnosis of the thermal
electron density and magnetic field strength along the line of sight.
The RM contribution from outside the Galaxy should be very small. The
effect of the EGS-intrinsic RM can only be accounted for by averaging
a sufficient large number of RMs in a small area. Unfortunately RM
data of EGSs are sparsely distributed in the sky despite recent
observational efforts. This holds in particular for the southern sky,
which makes the construction of an all-sky RM template a difficult
task. With the RM data compiled by \citet{hmq99} and the preliminary
RM data from the recent L-band Effelsberg RM survey of about 1800
polarized NVSS sources (Han et al., in prep.) we synthesized the RM
all-sky map shown in Fig.~\ref{effrm}, following the same
interpolation method as used by \citet{ufr+98} to create a smooth map
from ungridded data. The observed data close to each pixel of the
interpolated all-sky RM map are added with a weight calculated from
the function $\exp(-(\theta/\Theta)^{\rm m})$, where $\theta$ is the
angular distance to the pixel. $\Theta$ was set to $8\degr$ and m to
0.3. This RM map contains more data than used by \citet{dc05} and
\citet{jhe04}. However, the basic large-scale features seen in all RM
maps are rather similar. Although there are less data in
the southern sky, which causes edge effects as visible in Fig.~\ref{effrm}, 
the data clearly indicate the general RM trend as discussed below.  
We did not include the numerous RM data by \citet{btj03,bhg+07} from 
sources in the Galactic plane as the intention is to trace the 
high-latitude RM characteristics. We treat the Galactic plane RM 
characteristics separately.

\begin{figure}[!htbp]
\centering
\includegraphics[angle=-90,width=8.5cm]{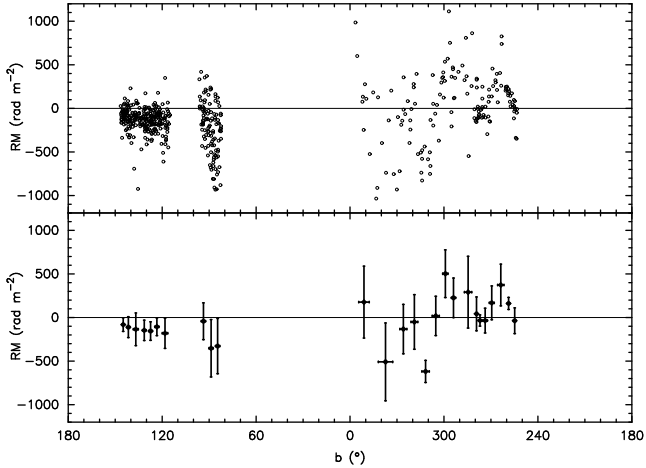}
\caption{RM profiles along the Galactic plane with RM data for the
northern Galactic plane taken from \citet{btj03} and RM data for the
southern Galactic plane from \citet{bhg+07}. The average for each
longitude bin is shown in the lower panel. The bars indicate the
$\pm1\sigma$ variance.}
\label{rmgp}
\end{figure}

The RM map in general shows an antisymmetric distribution below and
above the Galactic plane and also with respect to Galactic longitude
$0\degr$, as already noted by \citet{hmb+97}. In a recent RM study
towards the Galactic centre region this RM pattern was also seen
towards the inner Galaxy \citep{law}. The high-latitude RM values are
about 20--30~rad~m$^{-2}$ on average.

For the RM distribution along the Galactic plane we rely on the EGSs
detected in high-resolution Galactic plane surveys, where sources
contrast well against the Galactic diffuse emission. The Canadian
Galactic plane survey \citep[CGPS,~][]{tgp+03} covers parts of the
northern sky. The southern Galactic plane is covered by the SGPS
\citep{gdm+01}. The RM data from the CGPS \citep{btj03} and the SGPS
\citep{bhg+07} are shown in Fig.~\ref{rmgp}. Note that the fluctuation
of RMs is very large, typically around several hundred rad~m$^{-2}$,
and averaging over large areas is needed to trace systematic RM
changes. We sort the data into longitude bins and obtain their average
to be compared with the simulations. For the SGPS data the binning
procedure is the same as used by \citet{bhg+07}. For the CGPS data,
the RMs between $l=82\degr$ and $l=96\degr$ are divided into three
bins containing 38 or 39 sources and the RMs elsewhere are put into 11
bins containing 37 or 38 sources. The binned data are also shown in
Fig.~\ref{rmgp}. We note the following RM structural features: (1) the
average RMs run smoothly from about $-$100~rad~m$^{-2}$ at
$l\approx140\degr$ to about $-$200~rad~m$^{-2}$ at $l\approx100\degr$;
(2) the sign of the RMs changes at $l\approx304\degr$.  We believe
these features to be real. The local minimum of RMs at
$l\approx310\degr$ seems also reliable. There are some more features,
which seem less certain like RM values close to zero appearing at
$l\approx320\degr$ and $l\approx275\degr$. The RM fluctuations in the
Galactic centre region are extremely large \citep{rrs05}, which seem
to reflect strong variations of the properties of the interstellar
medium located in the centre region.  The available amount of RM data
is insufficient to allow to constrain the large-scale properties of
the interstellar medium and the magnetic field in the central region
of the Galaxy.

The large RM fluctuations observed might originate from the intrinsic
RMs of the EGSs or the interstellar medium in the Galaxy. Fluctuations
of the Galactic magnetic field have been estimated to be about
6~$\mu$G \citep{hfm04}, which might cause the large scatter of
RMs. However, we will show below that the clumpiness of thermal
electrons and a likely correlation with the random magnetic field
component causes large enough RM fluctuations for depolarization in
accordance with the observations. We note that strong RM fluctuations
are a handicap for the detection of cosmological magnetic fields from
fluctuations of foreground corrected RMs as reviewed by \citet{wid02}.

Currently the RMs of pulsars are not considered for modelling. First
of all the distances of pulsars are quite uncertain as many of them
are obtained from the dispersion measure based on global electron
density models, e.g. NE2001. Second, our simulation results are
all-sky maps, which are difficult to compare with the RMs of pulsars,
which are distributed along the line of sight. Finally, despite of
recent efforts, pulsar RM data are still sparsely distributed in the
Galactic plane and it is difficult to account for the influence of the
local interstellar medium. In future the pulsar RM data definitely
will be used as important additional constraints in particular as they
have no intrinsic RM contribution.

\begin{figure}[!htbp]
\includegraphics[angle=-90,width=9cm]{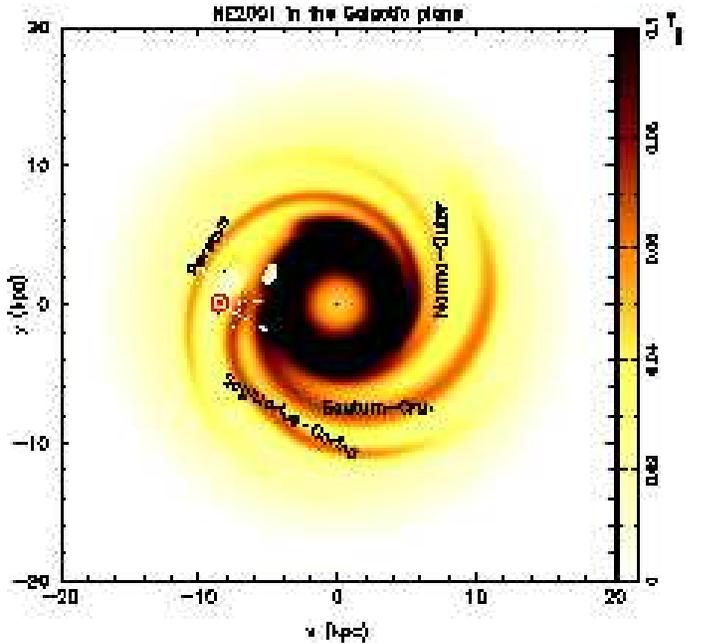}
\caption{The NE2001 thermal electron density model by
\citet{cl02,cl03} projected on the Galactic plane. The position of the
Sun as well as the designation of the spiral arms are indicated.}
\label{negp}
\end{figure}

\section{Numerical realization of Galactic 3D-modelling}

\citet{wae05} developed the HAMMURABI code aiming to simulate the
all-sky synchrotron emission of the Galaxy. The code takes Faraday
rotation effects into account and thus Faraday depolarization effects
are naturally involved. In this work we implemented additional entries
to account for optically thick free-free emission.  The code output
maps follow the HEALPIX pixelization scheme \citep{ghb+05}, which
divides the sky into pixels of equal areas. The total number of pixels
for an all-sky map, which defines its angular resolution, is $N_{\rm
pix}=12N_{\rm SIDE}^2$, with $N_{\rm SIDE}=2^k$ and
$k=0,1,2,3,\ldots$. The angular size of a pixel ($\Delta\theta$) can
be approximated as,
\begin{equation}
\Delta\theta\approx\sqrt{\frac{3}{\pi}}\frac{3600\arcmin}{N_{\rm SIDE}}
\end{equation}
In our simulation we use $N_{\rm SIDE}=256$, which corresponds to an
angular resolution of 15$\arcmin$. This selection is a compromise
between angular resolution and computing time.

Along the line of sight the cone for each pixel is divided into volume
elements in radial direction, where in this work each unit has an
equal radial length $\Delta r=20{\rm pc}$. The largest volume unit in
such a cone is approximately
\begin{equation} \label{eq::Vmax}
V_{\rm max}=4\pi r_{\rm max}^2 \frac{\Delta r}{12\cdot N_{\rm side}^2},
\end{equation} 
where $r_{\rm max}$ is the longest possible line of sight with the
maximal value of 32~kpc. Due to the disc like shape of our galaxy the
effective distance out to which we perform our computations varies and
decreases sharply towards high latitudes and towards the anti-centre
direction. Note that fluctuations of the magnetic fields on scales
smaller than the volume unit cannot be resolved by the code.

For the $i$-th volume element of a cone we compute the following
quantities,
\begin{equation}
\label{simequation}
\left\{
\begin{array}{rcl}
I_i^{\rm syn} &=& C_I B_{i,\,\bot}^{(1-p)/2} \nu^{(1+p)/2}\Delta r\\[2mm]
PI_i      &=& C_{PI} B_{i,\,\bot}^{(1-p)/2} \nu^{(1+p)/2}\Delta r\\[2mm]
RM_i          &=& 0.81n_{e\,i}B_{i,\,\parallel}\Delta r\\[2mm]
\psi_i        &=& \psi_{i,\,0}+\sum_1^iRM_j\lambda^2\\[2mm]
U_i           &=& PI_i\sin(2\psi_i)\\[2mm]
Q_i           &=& PI_i\cos(2\psi_i)\\[2mm]
EM_i          &=& n_{e\,i}^2 \Delta r\\[2mm]
\tau_i        &=& 8.235\times10^{-2} T_i^{-1.35} \nu^{-2.1} EM_i \\[2mm]
I_i^{\rm ff}  &=& T_i(1-\exp(-\tau_i))
\end{array}
\right.
\end{equation}

Note that the calculated Stokes values $U_i$ and $Q_i$ for each volume
element include the effect of foreground RM. The CR electrons are
assumed to follow a power law with an energy spectral index $p$.  The
values $C_I$ and $C_{PI}$ depend on the spectral index $p$ and are
further detailed in \citet{rl79}. The synchrotron emission is related
to the magnetic field component perpendicular to the line of sight
$B_{i,\,\bot}$, while the RM depends on the magnetic field parallel to
the line of sight $B_{i,\,\parallel}$. The intrinsic polarization
angle in each volume element is $\psi_{i,\,0}$, which is defined as
the inclination angle of $B_{i,\,\bot}$ with respect to the north (in
the frame of Galactic coordinates). Note that the magnetic field is
the sum of the regular and random field components. The thermal
electron density is $n_{e\,i}$ and its temperature is $T_i$. For the
relation between opacity $\tau_i$ and emission measure, EM, we refer
to \citet{rw00}. The calculation of $EM_i$, $\tau_i$ and $I_i^{\rm
ff}$ are expansions of the original HAMMURABI code.

Intensities and RMs for each pixel are obtained as the integral
of the contributions from all the volume units along the line of sight
\begin{equation}
\left\{
\begin{array}{ccc}
I^{\rm syn} & = & \sum_iI_i^{\rm syn}\\[2mm]
I^{\rm ff}  & = & \sum_iI_i^{\rm ff}\\[2mm]
Q           & = & \sum_iQ_i\\[2mm]
U           & = & \sum_iU_i\\[2mm]
RM          & = & \sum_iRM_i
\end{array}
\right.
\end{equation} 

All intensities are calculated as a function of frequency. More
details on the HAMMURABI code will be discussed in a forthcoming paper
by Waelkens, En{\ss}lin, et al.  (in prep.).

\begin{figure*}[!htbp]
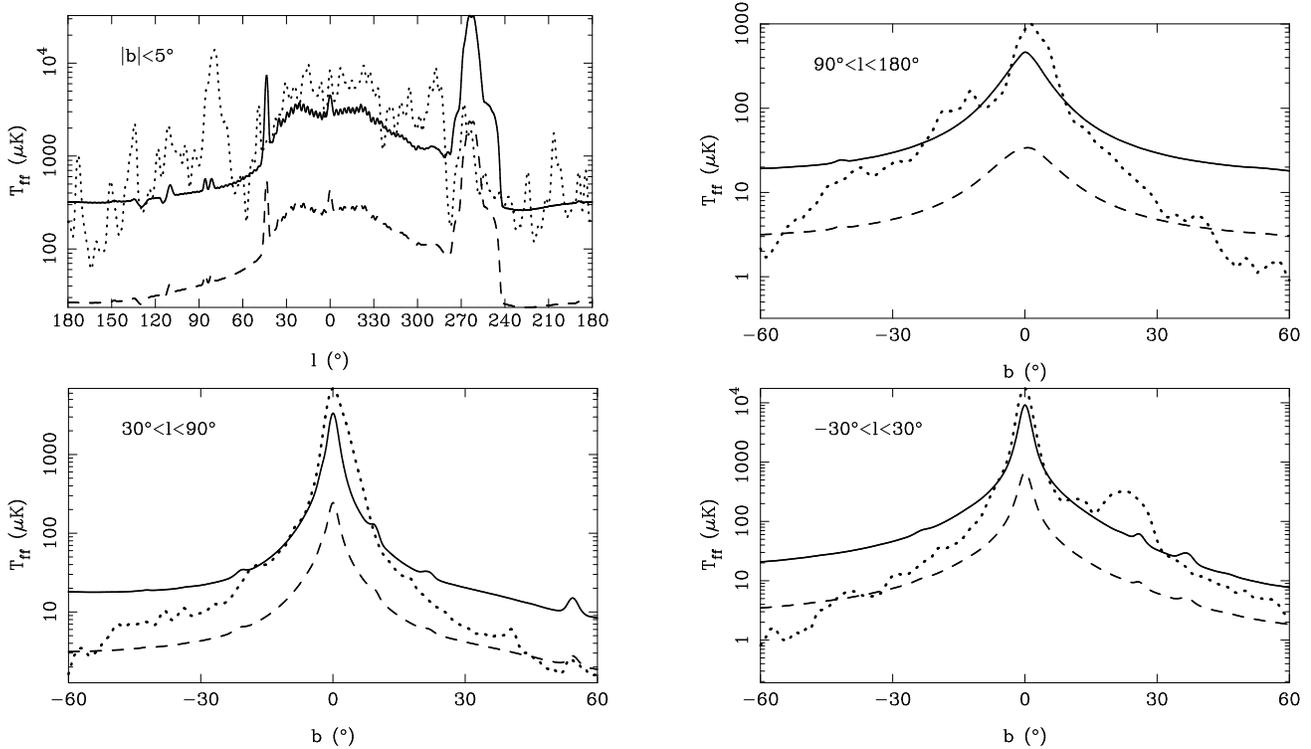

\centering
\begin{minipage}{0.5\textwidth}
\includegraphics[angle=-90,width=8cm]{ff_b_0.ps}
\end{minipage}%
\begin{minipage}{0.5\textwidth}
\includegraphics[angle=-90,width=8cm]{ff.l.90.180.ps}
\end{minipage}
\begin{minipage}{0.5\textwidth}
\includegraphics[angle=-90,width=8cm]{ff.l.30.90.ps}
\end{minipage}%
\begin{minipage}{0.5\textwidth}
\includegraphics[angle=-90,width=8cm]{ff.l.30.330.ps}
\end{minipage}
\caption{Longitude and latitude profiles from
the simulated and the free-free emission maps at 22.8~GHz.
The dotted lines indicate the data from the 22.8~GHz template. 
The solid and dashed lines are from the thermal emission simulations 
based on the NE2001 model without (dashed) and 
with (solid) a filling factor correction
according to \citet{bmm06}.}
\label{ffgp}
\end{figure*}

\section{Modelling}

For the description of the 3D-models we introduce cylindrical
coordinates ($R$, $\phi$, $z$), where $R$ is the Galactocentric
radius, $\phi$ is the azimuth angle starting from $l=180\degr$ and
increasing in anti-clockwise direction, and $z$ is the distance to the
Galactic plane. To show the slices of the 3D models a Galactocentric
Cartesian coordinate is also used, where $z$ is the same as in the
cylindrical system, and the $x$--$y$ plane coincides to the Galactic
plane with $x$ pointing to $l=0\degr$ and $y$ to $l=90\degr$.  Through
all calculations the distance of the Sun to the Galactic centre is
taken as 8.5~kpc.

Below we present a detailed 3D modelling of the Galactic emission
following the steps described in Sect.~3.

\subsection{Thermal electron density}

\subsubsection{The model for diffuse ionized gas -- NE2001}

According to \citet{fer01} the warm ionized medium (WIM) has an
electron density of 0.2 cm$^{-3}$ to 0.5 cm$^{-3}$, a volume filling
factor of 5$\%$ to 14$\%$ and a temperature of about 8000~K. It
consists of HII regions embedded in diffuse ionized gas (DIG). The
average electron density of HII regions can be derived from its radio
continuum integrated flux density \citep[e.g. ][]{sm69}. However, the
distances of HII regions are required, which are for instance derived
from observations of Doppler shifted recombination lines. Kinematic
distances are based on Galactic rotation curves and are not quite
certain. The scale height for HII regions is about 50~pc in the inner
Galaxy and increases towards larger radii \citep{pdd04}.

HII regions along the line of sight can influence RMs from sources in
the background as shown by \citet{mwkj03}. Since HII regions are
tightly confined to the Galactic plane, the chance that a line of
sight passes an HII region outside the plane is small. In the Galactic
plane the typical effect of HII regions on RMs is difficult to
estimate in general. Both the electron density and the magnetic field
inside a HII region may be an order of magnitude larger compared to
the diffuse gas. A typical RM contribution of several tens of rad
m$^{-2}$ was quoted by \citet{mwkj03}. These values are smaller
compared to the average RM of EGs, which trace the entire magnetized
DIG in the Galaxy along the line of sight. Therefore we consider only
the effect of the DIG in our simulations.

A good tracer of the DIG is the dispersion measure (DM) of
pulsars. The DM is defined as the integral of the electron density
along the line of sight. Using pulsars with known distances as
determined from e.g. parallax or HI-absorption measurements an
electron density model can be obtained by a fit of the observed
DMs. In this way a two-disk model was proposed and developed in detail
by \citet{cwf+91,tc93,gbc01}. \citet{cl02} and \citet{cl03} used more
pulsars and improved the model by involving local irregularities and
in particular added a spiral arm structure. Thus the NE2001 model has
three components: a thick disk, a thin disk and spiral arms as
displayed in Fig.~\ref{negp}. Locally the mid-plane electron density
is about 0.03~cm$^{-3}$ and the exponential scale height is about
0.97~kpc.

\subsubsection{The filling factor of the DIG -- from free-free emission at 
22.8~GHz}

The properties of the DIG can also be traced by its free-free
emission. This is done either by measuring H$\alpha$ intensities or
radio continuum emission. However, the diffuse Galactic thermal and
non-thermal radio emission components mix and are difficult to
separate. H$\alpha$ observations need an extinction correction to be a
reliable free-free emission template \citep{ddd03}, which is, however,
only possible out of the plane. Because the distances of the emission
components along the line of sight are unknown, the electron density
distribution cannot be obtained directly. However, the free-free
emission template can be used as a check to prove that the
integrated properties of the DIG model are appropriate.

Before we calculate the all-sky free-free emission map from the NE2001
model we assess the contributions from HII regions. The thermal
emission from HII regions is confined to a thin layer 
($|b|<1\degr$ e.g. \citet{pdd04}) along the Galactic plane with a
small scale height. We do not include individual HII regions in our
simulations of the free-free emission, which we believe is an
acceptable simplification.

The electron temperature is required to calculate the free-free
emission intensity as seen from Eq.~(\ref{simequation}). The existence
of an electron temperature gradient in the Galaxy is known and was
recently confirmed with some precision by \citet{qrb+06} from
recombination line measurements as $T_i(R)=(5780\pm350)+(287\pm46)R$,
where $R$ is in units of kpc. This corresponds to an electron
temperature of about 8000~K near the solar system, which is larger
than that derived by \citet{pdd04}. Note that these temperatures are
obtained from observations of HII regions, which are known to be lower
than the DIG by typically about 1000--2000~K \citep{mrh06}. However,
these temperature differences have just a small influence on the
amount of free-free emission (see Eq.  (\ref{simequation})). For
simplicity we use the temperature gradient established from the
measurements of HII regions.

\citet{rht99} reported a temperature increase with increasing
$z$. This increase was fitted by \citet{pw02} as
$T_i(z)=7000-526z+1770z^2$, where $z$ is in units of kpc. Combining
both temperature gradients we obtain
$T_i(R,z)=(5780\pm350)+(287\pm46)R-526z+1770z^2$ being used in our
calculation. Although the temperature is high at large $z$ the
free-free emission remains small due to the low electron density
there. Our results are not sensitive to the temperature gradient in
$z$. We note that the electron gradient in $z$ direction is still
debated. Recent results from \cite{ddb+06} based on
WMAP and ancillary data indicate rather low electron temperatures
 at high latitudes.

\begin{figure}[!htbp]
\centering
\includegraphics[width=9cm]{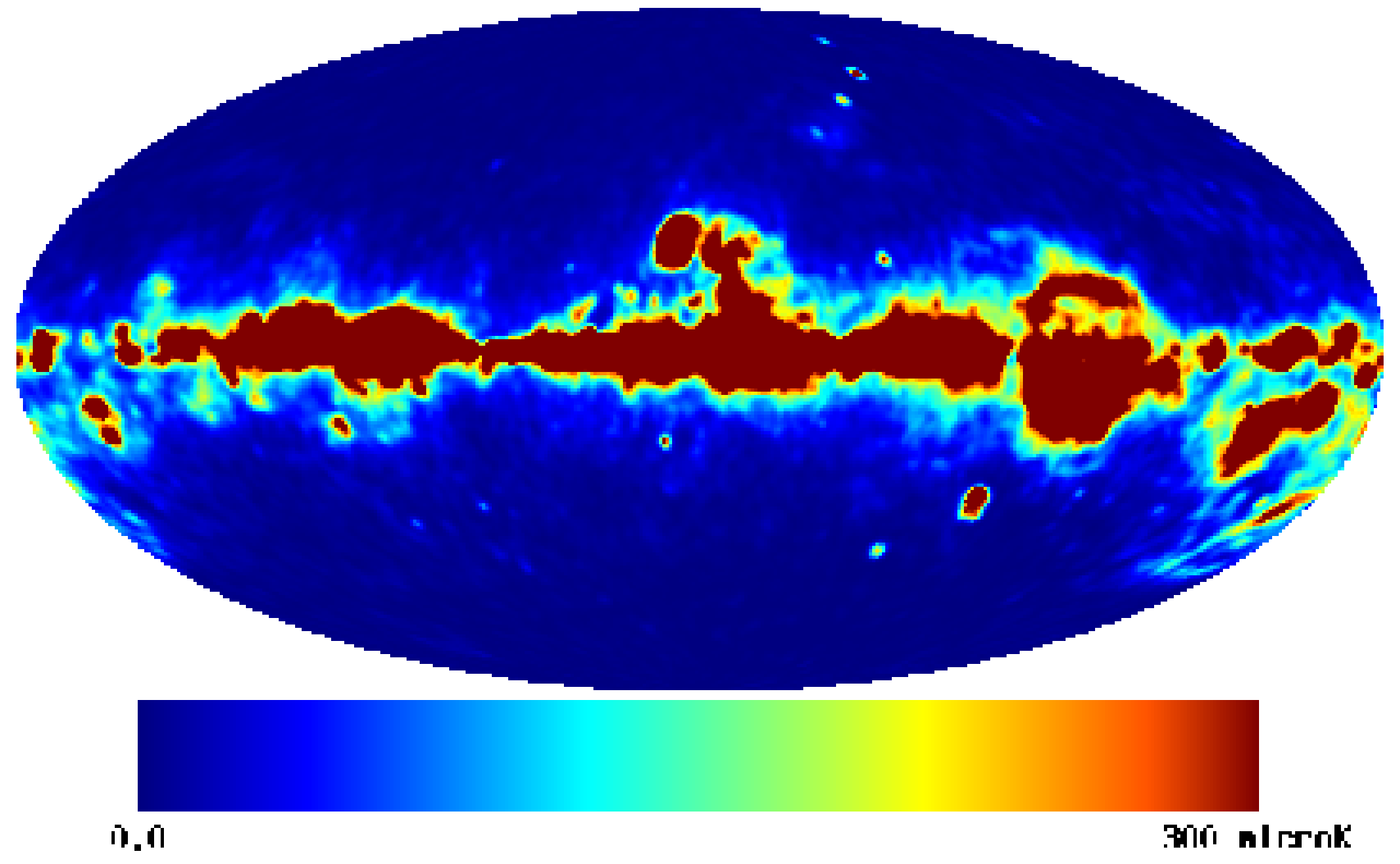}
\includegraphics[width=9cm]{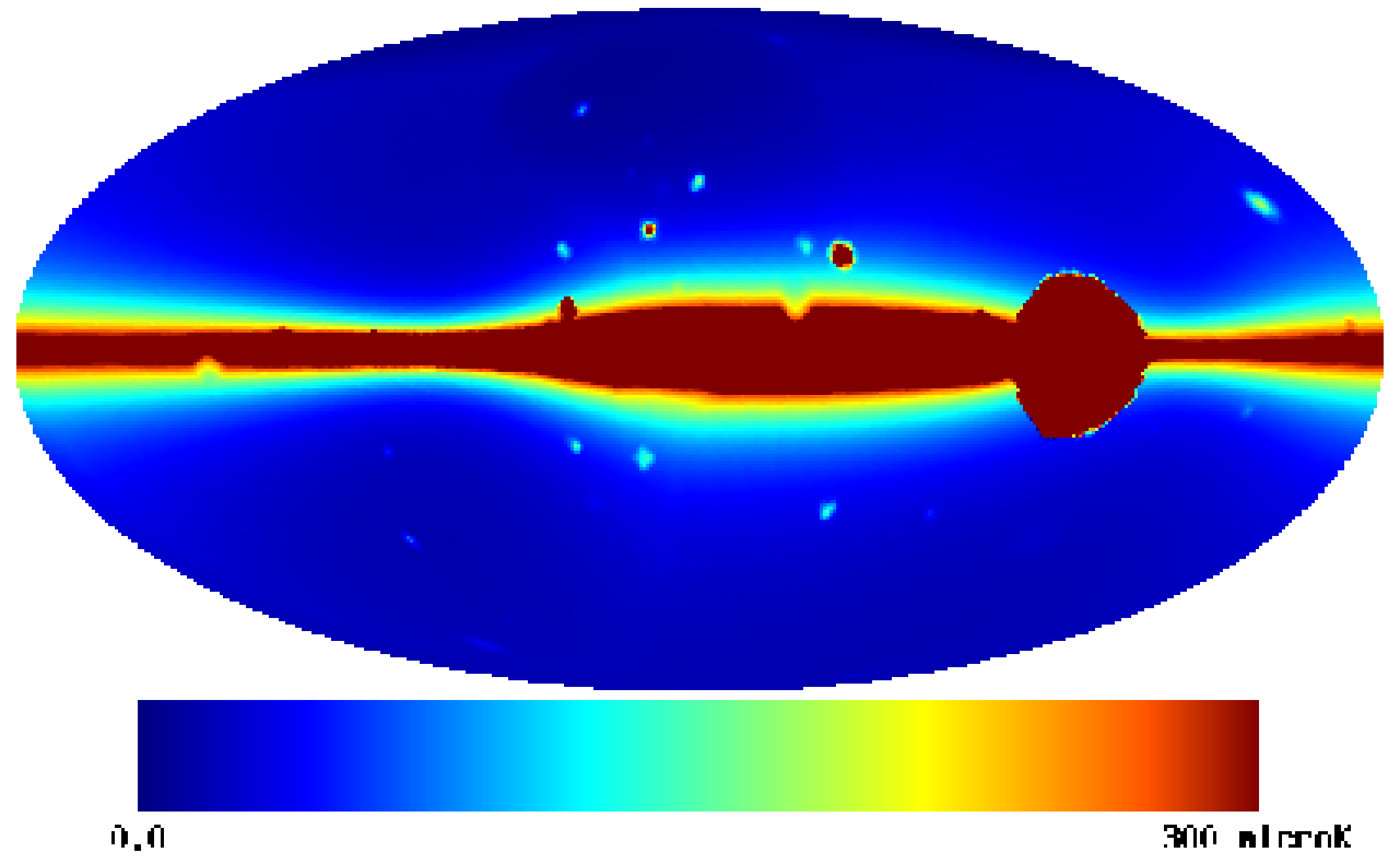}
\caption{The WMAP free-free emission template \citep{hnb+07} is shown
in the upper panel and the NE2001 based simulated maps after taking a
filling factor into account is shown in the lower panel. Both maps are
shown at an angular resolution of 2$\degr$.}
\label{ffmap}
\end{figure}

When we calculate the free-free emission from the NE2001 model as
shown in Fig.~\ref{negp} by using Eq.~(\ref{simequation}), we obtain
much smaller intensities than from the optically thin free-free
emission template derived by \citet{hnb+07} based on WMAP data. 
This can be seen from Fig.~\ref{ffgp} where the longitude profiles
(average for $|b|<5\degr$) and the latitude profiles (average for the
longitude ranges indicated in the plots) are shown. However, the
differences vanish when the clumpy nature of the DIG is taken properly
into account, which means to introduce an appropriate filling factor
of the thermal electrons. Recently \citet{bmm06} studied the Galactic
filling factor $f_e$ of the DIG and its dependence on $z$ in
detail. The filling factor $f_e$ is defined as $f_e=<n_e>^2/<n_e^2>$,
where the $<>$ means the average.  \citet{bmm06} derived
$f_e(z)=0.07\exp(|z|/0.5)$, where $z$ is in units of kpc, and
$f_e=0.32$ for $z$ larger than 0.75~kpc. EM calculated from the NE2001
model is based on $<n_e>$ and needs to be corrected by the factor
$1/f_e$, which means that EM increases by a factor $1/f_e$. After
employing the filling factor correction the comparison between the
free-free emission template from WMAP observations and the corrected
NE2001 model is shown in Figs.~\ref{ffgp} and \ref{ffmap}. The
simulations now resemble the WMAP template quite well. From the
comparison shown in Fig.~\ref{ffgp} it is also obvious that the
thermal high-latitude emission from the simulations is slightly higher
than that from the template, which might indicate a further increase
of the filling factor towards larger $z$ beyond the limit of 0.32
assumed by \citet{bmm06}.

\subsubsection{The filling factor of the DIG -- from low-frequency absorption}

\begin{figure}[!htbp]
\centering
\includegraphics[angle=-90,width=8.5cm]{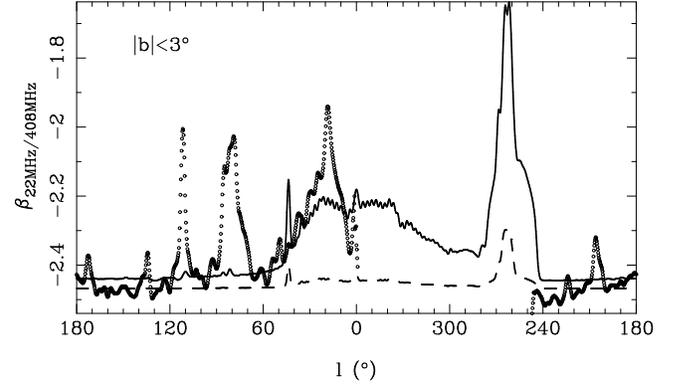}
\caption{The average spectral index along the Galactic plane
($|b|<3\degr$) calculated from two simulated maps at 408~MHz and
22~MHz including thermal absorption from the NE2001 model with (solid
line) and without a filling factor correction (dashed line). For
comparison we show the observed spectral indices ($|b|<3\degr$)
calculated from the 22~MHz northern sky survey map \citep{rcls99} and
the 408~MHz map \citep{hssw82} at an angular resolution of $2\degr$
(circles). }
\label{spec_absorp}
\end{figure}

At low frequencies the optical thickness of free-free emission
increases (Eq. \ref{simequation}) which implies that the synchrotron
emission from behind is attenuated by a factor exp$(-\tau)$. This can
be seen by an apparent increase of the spectral index by
$\tau/\ln(\nu_h/\nu_l)$, where $\nu_h$ and $\nu_l$ are the high and
the low frequency, respectively. Since the optical thickness is
proportional to EM the filling factor correction is essential. We
simulate total intensity maps at 408~MHz and 22~MHz by using an
observed spectral index of $\beta=-2.47$ obtained by \citet{rcls99}
from the 408~MHz and 22~MHz surveys for regions outside the Galactic
plane, which closely represents the spectral index of the Galactic low
frequency synchrotron emission. The spectral index map is calculated
from the simulated maps at the two frequencies and the average profile
along the Galactic plane is compared with that observed
(Fig.~\ref{spec_absorp}). We find that the spectral index
increase from $l=60\degr$ towards the Galactic centre is well
reproduced by our simulation.  Note that the peaks are from individual
source complexes, which are not included in our model. For comparison
we show also the spectral index profile from the NE2001 model without
the filling factor correction, which does not show the spectral
increase towards the inner Galaxy caused by thermal absorption.

\subsubsection{Remarks}

The NE2001 model represents the average large-scale electron density
distribution, including some sources and large complexes like the Gum
nebula.  However, some similarly prominent features like the Cygnus
complex at about $l \sim 90\degr$ are missing as it is clearly seen in
the free-free emission template and the low-frequency absorption
spectra. This seems to be related to the much larger number of pulsars
in the southern sky, which allows a modelling of such features
there. After considering appropriate filling factors the NE2001 model
is able to reproduce the WMAP free-free emission template at 22.8~GHz
and also the low-frequency absorption quite well.

In summary the NE2001 model combined with the filling factor obtained
by \citet{bmm06} are proven to be sufficient to model the low latitude
DIG. We will discuss later that there are observational problems
suggesting that at high-latitudes the thermal electrons density is
underestimated.

\subsection{Regular magnetic field properties}

The Galactic magnetic field is commonly described as a vector with a
random component $\mathbf{b}$ and a regular component $\mathbf{B}$. It
can be written as, $\mathbf{B}_{\rm tot}=\mathbf{B}+\mathbf{b}$.
Following the procedure described in Sect.~3, the random field
component will be discussed together with the CR electron
distribution.  The regular field properties are constrained by
RMs. Below we use $B=|\mathbf{B}|$ to denote the strength of the
magnetic field vector. We treat the regular field component as the
combination of a disk field $\mathbf{B}^D$ and a halo field component
$\mathbf{B}^H$. The former is constrained by RMs near the Galactic
plane ($|b|<5\degr$) and the latter by an all-sky RM map from EGS
data.

\subsubsection{The disk field}

It is generally accepted that our Galaxy has an organized large-scale
disk field similar to other nearby Galaxies \citep[see reviews by
e.g.~][]{hw02}.  As described by \citet{bbm+96} the field more or less
follows a spiral pattern that can be basically classified into an
axi-symmetric spiral (ASS) with no dependence on the azimuthal angle
or a bi-symmetric spiral (BSS), which has a symmetry of
$\pi$. Relative to the Galactic plane the field above and below can be
either symmetric (even parity or quadrupole) or asymmetric (odd parity
or dipole).

\begin{figure*}[!htbp]
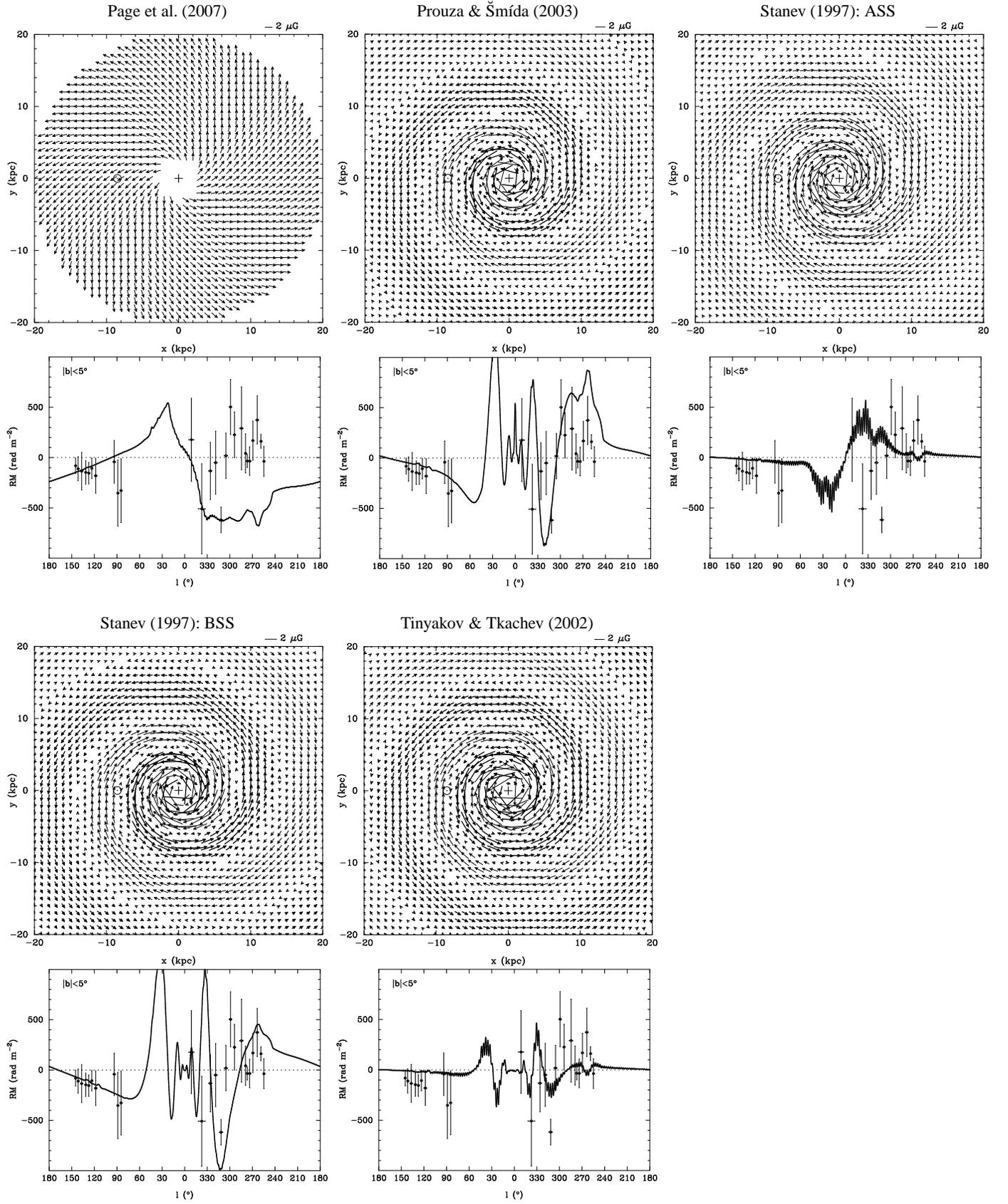

\begin{minipage}[t]{0.33\textwidth}
\centering
\citet{phk+07}
\end{minipage}
\begin{minipage}[t]{0.33\textwidth}
\centering
\citet{ps03}
\end{minipage}
\begin{minipage}[t]{0.33\textwidth}
\centering
\citet{sta97}: ASS
\end{minipage}
\begin{minipage}{0.33\textwidth}
\centering
\includegraphics[angle=-90,width=6cm]{b_lsa.ps}
\includegraphics[angle=-90,width=6cm]{rm_lsa.ps}
\end{minipage}
\begin{minipage}{0.33\textwidth}
\centering
\includegraphics[angle=-90,width=6cm]{b_ps.ps}
\includegraphics[angle=-90,width=6cm]{rm_ps.ps}
\end{minipage}
\begin{minipage}{0.33\textwidth}
\centering
\includegraphics[angle=-90,width=6cm]{b_sta_ass.ps}
\includegraphics[angle=-90,width=6cm]{rm_sta_ass.ps}
\end{minipage}\vspace{0.5cm}
\begin{minipage}[t]{0.33\textwidth}
\centering
\citet{sta97}: BSS
\end{minipage}
\begin{minipage}[t]{0.33\textwidth}
\centering
\citet{tt02}
\end{minipage}\\
\begin{minipage}{0.33\textwidth}
\centering
\includegraphics[angle=-90,width=6cm]{b_sta_bss.ps}
\includegraphics[angle=-90,width=6cm]{rm_sta_bss.ps}
\end{minipage}
\begin{minipage}{0.33\textwidth}
\centering
\includegraphics[angle=-90,width=6cm]{b_tt.ps}
\includegraphics[angle=-90,width=6cm]{rm_tt.ps}
\end{minipage}
\caption{Different magnetic field configurations in the Galactic plane
as seen from the north Galactic pole. The corresponding RM profiles
along the Galactic plane ($|b|<5\degr$) are shown in the panel
below. The average of RMs within longitude bins from the CGPS and the
SGPS are shown for comparison.}
\label{rmall}
\end{figure*}

\begin{figure*}[!htbp]
\begin{minipage}[t]{0.33\textwidth}
\centering
ASS+RING
\end{minipage}
\begin{minipage}[t]{0.33\textwidth}
\centering
ASS+ARM
\end{minipage}
\begin{minipage}[t]{0.33\textwidth}
\centering
BSS
\end{minipage}
\begin{minipage}{0.33\textwidth}
\includegraphics[angle=-90,width=6cm]{b_ring.ps}
\includegraphics[angle=-90,width=6cm]{rm_ring.ps}
\end{minipage}
\begin{minipage}{0.33\textwidth}
\includegraphics[angle=-90,width=6cm]{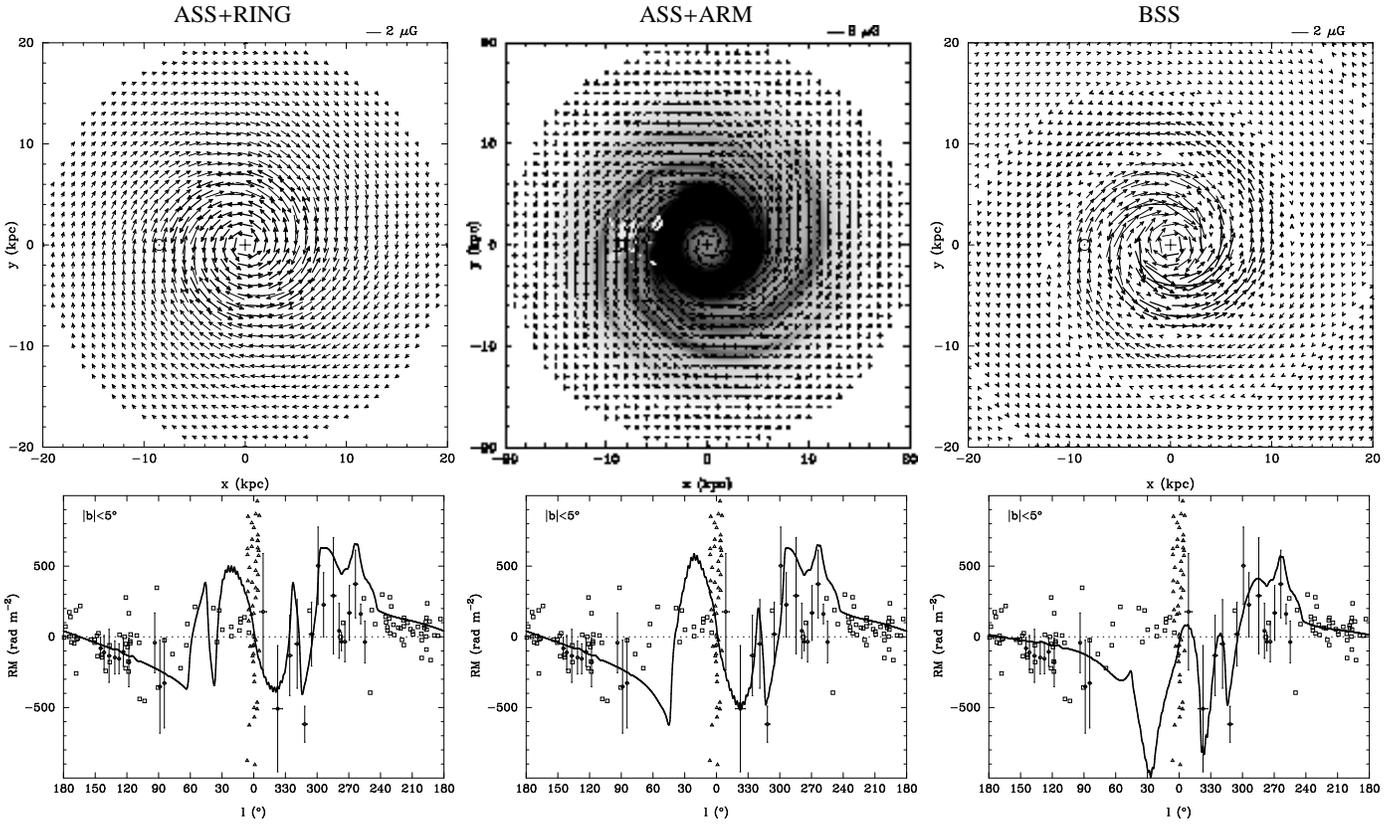}
\includegraphics[angle=-90,width=6cm]{rm_arm.ps}
\end{minipage}
\begin{minipage}{0.33\textwidth}
\includegraphics[angle=-90,width=6cm]{b_bss.ps}
\includegraphics[angle=-90,width=6cm]{rm_bss.ps}
\end{minipage}
\caption{The same as Fig.~\ref{rmall}, but for the ASS+RING, the
ASS+ARM and the BSS field configurations. Beside the binned CGPS and
SGPS data the RMs near the center from \citet{rrs05} and RMs from the
compilation of \citet{hmq99} are shown by triangles and squares,
respectively. For the ASS+ARM model the NE2001 model is overlaid onto
the field in gray scale.}
\label{rm_3}
\end{figure*}

Magnetic fields are often discussed in the context of propagation of
high-energy cosmic rays. The ASS field configuration has been used by
\citet{sta97} and the BSS field configuration by \citet{sta97},
\citet{tt02} and \citet{ps03} for modelling. A logarithmic spiral arm
field was proposed by \citet{phk+07} to interpret the WMAP
observations. We collected a number of published models and checked
them for their ability to simulate the observed RMs in the Galactic
plane. The results are shown in Fig.~\ref{rmall}. Obviously none of
the models is able to reproduce the observed systematic features. This
motivates us to revise the parameters previously used.

The general form of a coplanar and constant pitch angle spiral
magnetic field configuration can be written in the cylindrical
coordinate system as
\begin{equation}
\left\{
\begin{array}{rcl}
B^D_R    & = & D_1(R,\phi,z)D_2(R,\phi,z)\sin p\\[2mm]
B^D_\phi & = &-D_1(R,\phi,z)D_2(R,\phi,z)\cos p\\[2mm]
B^D_z    & = & 0 
\end{array}
\right.
\end{equation}
where $D_1(R,\phi,z)$ constrains the spatial variation of the field
strength and $D_2(R,\phi,z)$ introduces reversals or assymmetries. The
definition of the pitch angle $p$ is the same as previously used by
\citet{hq94}.  We use the function $D_1(R,\phi,z)$ always in the
following way:
\begin{equation}\label{ring}
D_1(R,z)=\left\{
\begin{array}{cl}
\displaystyle{B_0\exp\left(-\frac{R-R_\odot}{R_0}-\frac{|z|}{z_0}\right)} & 
R>R_c\\[4mm]
B_c & R\leq R_c
\end{array}
\right.
\end{equation}

However, for different models the variables $R_0,~R_c$ and $z_0$ may
change.  This kind of variation is the same as used by \citet{hml+06}
to describe the magnetic field from pulsar RM observations. Too few RM
data were observed so far in the first quadrant, which does not allow
a detailed modelling of the field within the inner Galaxy. Therefore
we keep the field strength constant there.

The regular $z$-component of the Galactic magnetic field is very
small. Its strength is about 0.2~$\mu$G to 0.3~$\mu$G as estimated by
\citet{hq94}. We assume $B_z^D$ to be zero in our models.

Below we list three different magnetic field models: (1) the ASS model
plus reversals in rings (ASS+RING); (2) the ASS model plus reversals
following arms (ASS+ARM); and (3) the BSS model. Following the
procedure described in Sect.~3 we use the NE2001 model for the thermal
electrons to adjust the parameters in the three models by a
trial-and-error approach in order to achieve the best fit for the
observed RMs in the plane. The results are summarized below:

{\it ASS+RING---} In this model, the parameters in $D_1(R,z)$ are
$R_0=10$~kpc, $z_0=1$~kpc, $R_c=5$~kpc, $B_0=2$~$\mu$G and
$B_c=2$~$\mu$G. The field reversals are specified by $D_2(R)$ being
symmetric in $z$ as,

\begin{equation}
D_2(R)=\left\{
\begin{array}{ll}
+1 & R>7.5\,\,{\rm kpc}\\
-1 & 6\,\,{\rm kpc}<R\leq 7.5\,\,{\rm kpc}\\
+1 & 5\,\,{\rm kpc}<R\leq 6\,\,{\rm kpc}\\
-1 & R\leq 5\,\,{\rm kpc} 
\end{array}
\right.
\end{equation}

Here $+1$ means a clockwise direction as seen from the north pole. The
pitch angle is taken as $-12\degr$, which is the average of pitch
angles used for the spiral arms in the NE2001 model.

It can be seen from Fig.~\ref{rm_3} that the simulated RMs in the
Galactic plane fit the data of the CGPS and the SGPS well.  We also
used the RM data from \citet{rrs05} near the centre and elsewhere from
the compilation of \citet{hmq99}. Although the number of measurements
is small they basically support the results from the simulations.

{\it ASS+ARM---} In this model the parameters for $D_1(R,z)$ are
$R_0=8.5$~kpc, $z_0=1$~kpc, $R_c=5.3$~kpc, $B_0=2$~$\mu$G and
$B_c=2$~$\mu$G. We use similar logarithmic arms as \citet{cl03} (see
Fig.~\ref{negp}), but with a pitch angle of $-12\degr$.  The pitch
angle of the magnetic field is the same as that of the arms.  The
reversals are placed between the inner edge of the Sagittarius-Carina
arm and the inner edge of the Scutum-Crux arm.  This reversal is
consistent with that found by \citet{bhg+07} based on the SGPS RM
data. However, this should not be taken as a final conclusion because
the amount of RM data is small and also the position and the extent of
the spiral arms is uncertain.  As can be seen from Fig.~\ref{rm_3}
this model fits the observations as well as the ASS+RING model.

{\it BSS---} In this model the parameters for $D_1(R,z)$ are
$R_0=6$~kpc, $z_0=1$~kpc, $R_c=3$~kpc, $B_0=2$~$\mu$G and
$B_c=2$~$\mu$G. $D_2(R,\phi)$ is written as
\begin{equation}
D_2(R,\phi)=-\cos\left(\phi+\beta\ln\frac{R}{R_b}\right)
\end{equation} 
where $\beta=1/\tan p$. The parameters are $R_b=9$~kpc and
$p=-10\degr$ for $R>6$~kpc and otherwise $R_b=6$~kpc and
$p=-15\degr$. The model is able to fit the RM data from the SGPS and
the CGPS (Fig.~\ref{rm_3}), but is inconsistent with other data as
discussed below.

\subsubsection{The halo field}

The synthesized all-sky RM map (Fig.~\ref{effrm}) shows an asymmetry
in longitude and latitude relative to the Galactic plane and the
Galactic centre, respectively, indicating an asymmetric halo magnetic
field configuration.

Such a RM configuration could in principle also originate from large
local shells.  However, at least two shells with opposite magnetic
field orientations located below and above the plane are needed.
\citet{hmq99} showed that the RMs of pulsars with distances larger
than 1~kpc increase with distance, which makes an explanation by local
shells unlikely and strongly supports the large-scale field scenario.

\begin{figure}[!htbp]
\includegraphics[angle=-90,width=9cm]{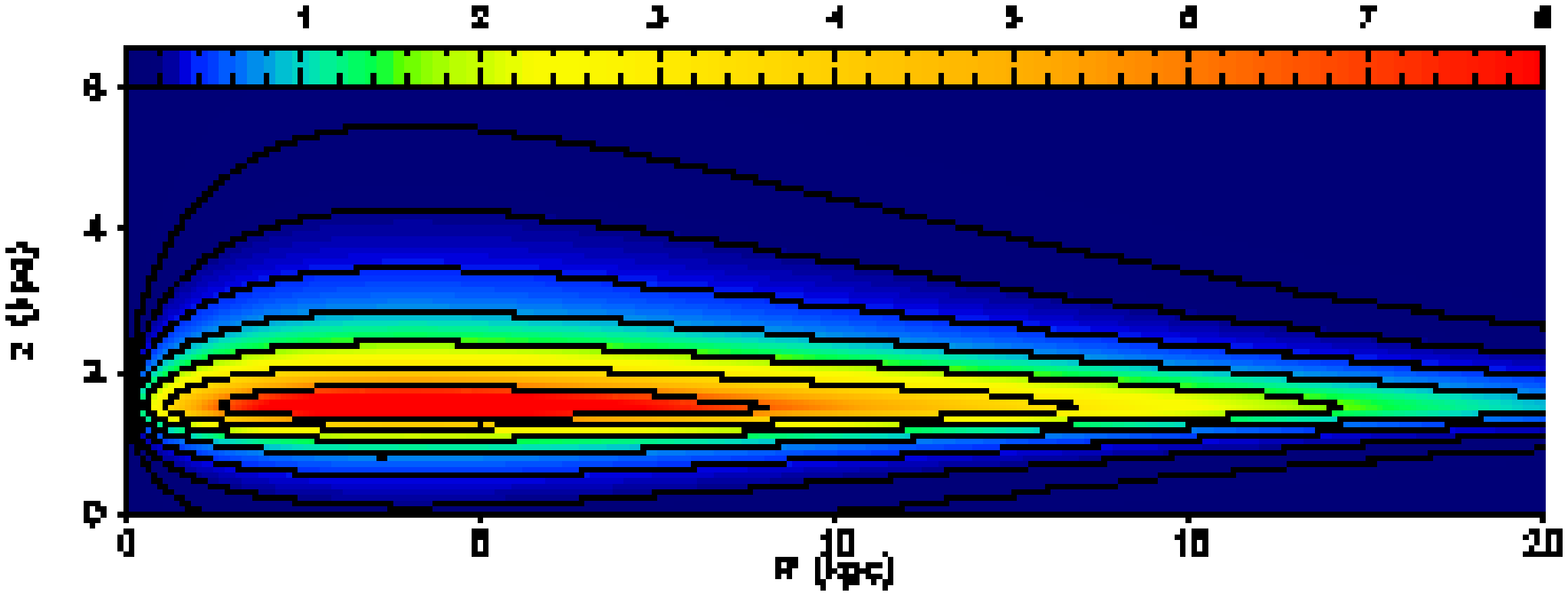}
\caption{The strength of the regular halo magnetic field component. 
Contours start at 0.1~$\mu$G and increase by a factor of 2.
The field below the plane has the same strength but is reversed.}
\label{bhalo}
\end{figure}

We follow \citet{ps03} for a description of the regular double-torus
field component for one half of the Galaxy and the reversed direction
in the other half:
\begin{equation}
B^H_\phi(R,z)=B^H_0\frac{1}{1+\left(\displaystyle{\frac{|z|-z^H_0}{z^H_1}}\right)^2}
\frac{R}{R^H_0}\exp\left(-\frac{R-R^H_0}{R^H_0}\right)
\end{equation}
The parameters are $z^H_0=1.5$~kpc, $z^H_1=0.2$~kpc for $|z|<z^H_0$
and otherwise $z^H_1=0.4$~kpc, $B^H_0=10$~$\mu$G, and $R^H_0=4$~kpc.
The strength of the halo magnetic field above the plane is shown in
Fig.~\ref{bhalo}.

\begin{figure}
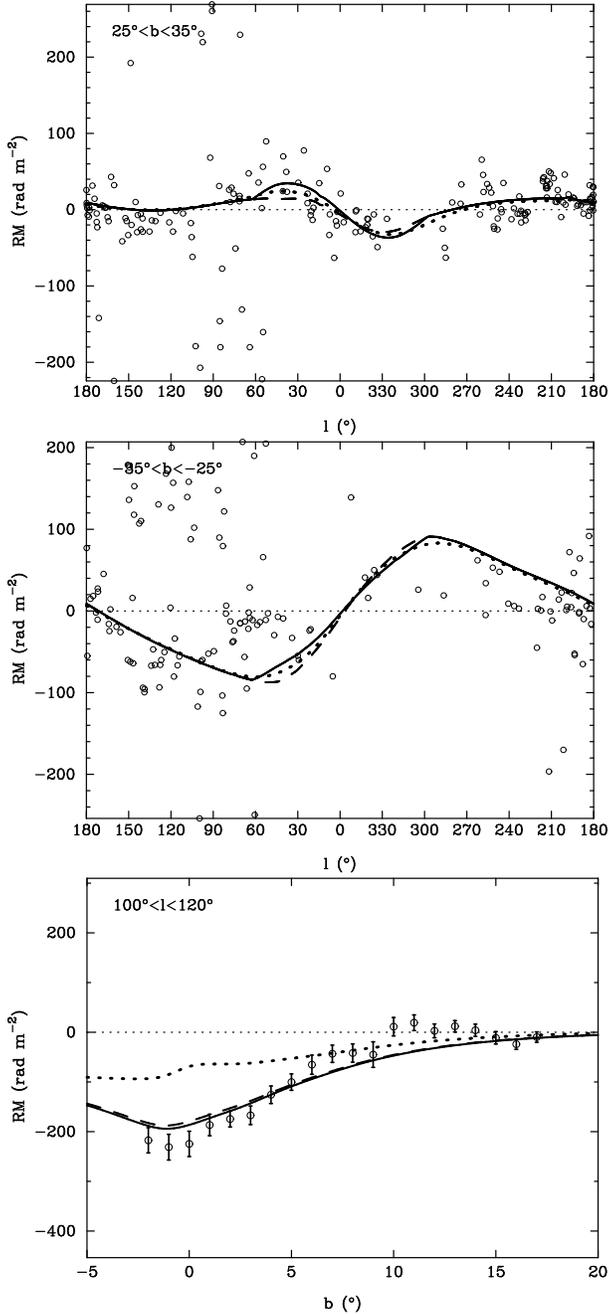

\includegraphics[angle=-90,width=8cm]{rmhaloa.ps}
\includegraphics[angle=-90,width=8cm]{rmhalob.ps}
\includegraphics[angle=-90,width=8cm]{rmvert.ps}
\caption{RM profiles from the halo field plus the ASS+RING model 
(solid), the ASS+ARM (dashed) model and the BSS model (dotted) 
versus the observed RM data. The RM data in the bottom panel
are from the extended CGPS averaged
for latitude intervals of $2\degr$ (see text).}
\label{rmhalo}
\end{figure}

We add the halo field to the disk field, run the simulation and obtain
average longitude profiles for $25\degr<b<35\degr$ and
$-35\degr<b<-25\degr$ as shown in Fig.~\ref{rmhalo}. Some RMs of EGSs
at longitudes of about 90$\degr$ and larger show large deviations both
above and below the plane. These areas coincide with the positions of
Loop II and Loop III. However, whether the Loop structures physically
cause these RM deviations needs a further check. Except for these
outliers the simulations are generally consistent with the
observational data.
 
We show the latitude variation of the simulated RMs in
Fig.~\ref{rmhalo} and in addition the average RM between $l=100\degr$
and $l=120\degr$ versus latitude from $-5\degr$ to $20\degr$.  The
observed data are from the extended CGPS (Jo-Anne Brown et al., in
prep). The halo field plus the ASS+RING or the ASS+ARM disk fields
yield quite a good fit.  However, the halo field plus the BSS field is
clearly inconsistent with the data. If we tune the BSS field to match
the latitude profile, we fail to get a good fit to the RMs along the
Galactic plane. We conclude that the BSS model is not in agreement
with the observed RMs of EGSs.
 
\subsubsection{Remarks}

We introduced field reversals in our models with sharp transitions,
which means an abrupt field change in the opposite direction. This
might be not a realistic scenario, however, we do not expect much
influence by this simplification as the transition regions are
smoothed out in our simulated maps. We also used a constant pitch
angle in our models, which is most likely another simplicifation. We
note that the ASS+RING and the ASS+ARM models are not very sensitive
to the pitch angle used.

\subsection{Random fields, CR electron density and local excess of the 
synchrotron emission}

\subsubsection{The random magnetic field component}

The knowledge about the properties of random magnetic fields in
galaxies is quite limited \citep{b01}. A typical scale for the random
fields was claimed to be about 50~pc based on the single cell size
model \citep{rk89}. However, the single cell model is a strong
simplification. Structure functions of RMs of EGSs yield the
fluctuation properties of $n_eb_\parallel$. Unfortunately, due to the
coupling between the electron density and magnetic field the
extracting of the magnetic field fluctuation remains very
uncertain. \citet{ms96} tried to tackle the decoupling by involving
the structure function of emission measures and concluded that the
magnetic field spectrum can be described by a Kolmogorov spectrum with
an outer scale of about 4~pc. The turbulence for the region of sky
investigated by \citet{ms96} occurs in thin sheets or filaments and
can be described as a two-dimensional turbulence for scales between
4~pc to about 80~pc. \citet{hfm04} claimed a flatter spectrum for
scales between 500~pc and 1.5~kpc based on pulsar RM and DM data.

In principle a stochastic realisation of the random magnetic fields
can be generated by the Fourier transform of the square root of the
spectrum as \citep[e.g. ][]{mar05}
$b(\mathbf{s})\propto\int\sqrt{P(\mathbf{k})}(\xi+i\zeta)
\exp(i\mathbf{k}\cdot\mathbf{s}){\rm d}\mathbf{s}$, where
$P(\mathbf{k})$ is the power spectrum, $\mathbf{k}$ is the wavenumber,
$\mathbf{s}$ is the spatial position, and $\xi$ and $\zeta$ are
Gaussian random numbers. This requires a 3D inverse Fourier
transformation to obtain $b(\mathbf{s})$. To fulfill the above
equation the Galaxy must be gridded to small cubic boxes with a volume
of $l^3$. To avoid an intrinsic correlation by the selected box size,
they must be smaller than the outer scale of the magnetic field
fluctuations. Let $l$ have a size of 2~pc and the Galaxy volume to be
40~kpc$\times$40~kpc$\times$10~kpc $2\times10^{12}$ boxes for an
all-sky calculation are needed, which is far beyond current computer
memory capabilities. However, a small patch of the sky can be treated
that way, utilizing a huge amount of computing time. A more detailed 
discussion on how to obtain a random field from a Kolmogorov spectrum for
high-angular resolution simulations will be presented in a subsequent
paper, where we will simulate high-resolution patches of Galactic
emission as they will be observed by LOFAR and the future SKA.

Here we assume that the random field components are homogeneous and
follow a Gaussian distribution in strength with an average of zero and
a scatter of $\sigma_b/\sqrt{3}$ in each dimension.  As discussed
later the best value for the mean random field strength is $\mathbf{b}
= 3~\mu$G throughout the Galaxy.

\subsubsection{CR electron density}

Synchrotron emission in the frequency range from 10~MHz to 40~GHz
originates from CR electrons with energies between 400~MeV and 25~GeV
assuming a magnetic field component perpendicular to the line of sight
of 4~$\mu$G \citep{wsc80}. Local measurements of the density of CRs
with energies below several GeV are very uncertain due to the limited
knowledge of the heliospheric modulation \citep[e.g. ][]{smr04}. For
the CR electrons with energies higher than several GeV, the
measurements \citep[e.g.][]{bbb+98,gsc+02} follow a simple power law.
The flux density can be described as $N(\gamma)=C(R,z)\gamma^p$ with
$p$ of about $-$3, where $\gamma$ is the Lorentz factor. The CR flux
density at 10~GeV from the aforementioned measurements is about
0.2~(GeV~m$^2$~sr~s)$^{-1}$ which means $C(R=8.5\,\,{\rm
kpc},z=0)\equiv C_0\approx3.2\times10^{-5}$~cm$^{-3}$.  However, it is
questionable wheather the local measurements are representative for
the CR density elsewhere in the Galaxy \citep{pe98,smr04}. Therefore
we regard below the measured local CR flux density as weakly constrained.

We assume that the CR electron distribution consists of a disk component,
whose spatial distribution is written as
\begin{equation}
C(R,z)=C_0\exp\left(-\frac{R-R_\odot}{8\,\,{\rm kpc}}-\frac{|z|}
{1\,\,{\rm kpc}}\right)
\end{equation}
with $C_0=6.4\times10^{-5}$~cm$^{-3}$. For the Galactic
centre region the CR electron flux density is taken as a constant
$C(R<3\,\,{\rm kpc})=C(R=3\,\,{\rm kpc})$. The disk component is
set to zero for $z>1$~kpc as discussed in Sect.~7.2.

The spectral index of the CR electrons $p$, the radio flux density
$\alpha$ and the brightness temperature $\beta$ are related to each
other as $\alpha=(p+1)/2=\beta+2$. Spectral index maps show spatial
variations of the spectral index and a spectral steepening at high
frequencies is indicated \citep{rrt04}.  In this paper we simplify the
problem and keep the energy spectral index of the CR electrons $p$
constant with a value of $p =-3$ corresponding to $\beta=-3$. 
For the frequency range below 408~MHz we assume $p$ to be $-2$ or 
$\beta=-2.5$.

\begin{figure*}[!htbp]
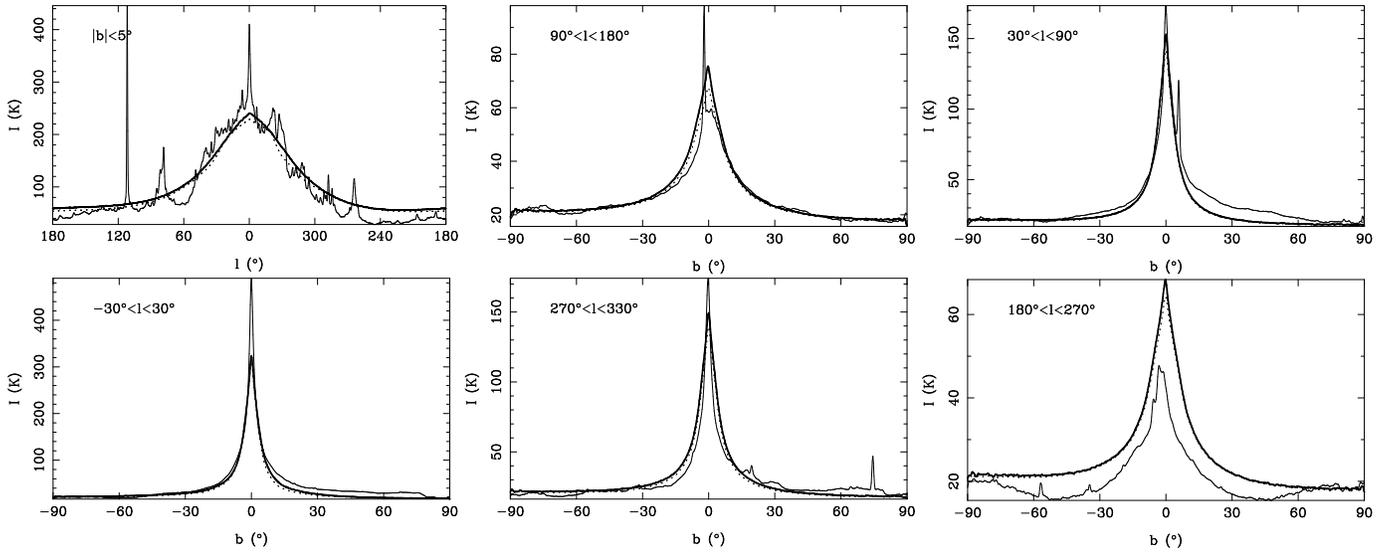

\begin{minipage}{\textwidth}
\includegraphics[angle=-90,width=6cm]{i.408.obs.sim.b.0.ps}
\includegraphics[angle=-90,width=6cm]{i.408.obs.sim.l.180.90.ps}
\includegraphics[angle=-90,width=6cm]{i.408.obs.sim.l.90.30.ps}
\end{minipage}
\begin{minipage}{\textwidth}
\includegraphics[angle=-90,width=6cm]{i.408.obs.sim.l.30.330.ps}
\includegraphics[angle=-90,width=6cm]{i.408.obs.sim.l.330.270.ps}
\includegraphics[angle=-90,width=6cm]{i.408.obs.sim.l.270.180.ps}
\end{minipage}
\caption{Total intensity profiles at 408~MHz for three magnetic field
models. The thin solid lines stand for the profiles from the 408~MHz
all sky survey. The results from the ASS+RING (thick solid lines), the
ASS+ARM (dashed) and the BSS (dotted) disk field models are also
shown. The simulated maps have been smoothed to an angular resolution
of 51$\arcmin$.}
\label{toti}
\end{figure*}
 
\begin{figure*}[!htbp]
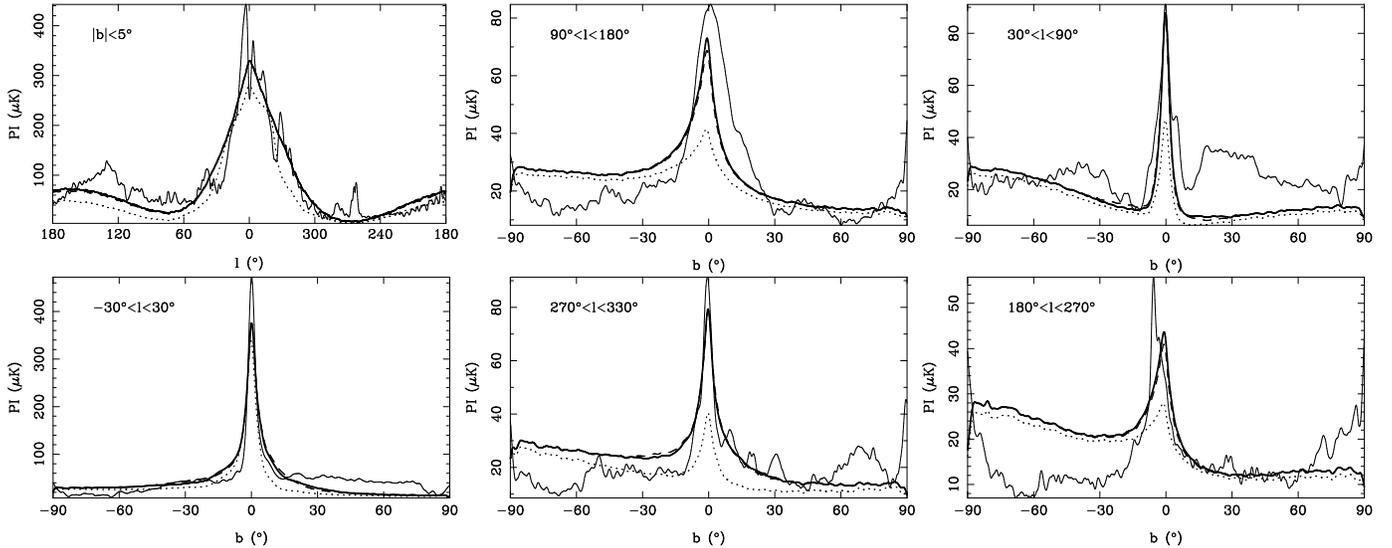

\begin{minipage}{\textwidth}
\includegraphics[angle=-90,width=6cm]{pi.22.8.obs.sim.b.0.ps}
\includegraphics[angle=-90,width=6cm]{pi.22.8.obs.sim.l.180.90.ps}
\includegraphics[angle=-90,width=6cm]{pi.22.8.obs.sim.l.90.30.ps}
\end{minipage}
\begin{minipage}{\textwidth}
\includegraphics[angle=-90,width=6cm]{pi.22.8.obs.sim.l.30.330.ps}
\includegraphics[angle=-90,width=6cm]{pi.22.8.obs.sim.l.330.270.ps}
\includegraphics[angle=-90,width=6cm]{pi.22.8.obs.sim.l.270.180.ps}
\end{minipage}
\caption{The same as Fig.~\ref{toti}, but for profiles of polarized
intensity at 22.8~GHz. All the maps are smoothed to an angular
resolution of 2$\degr$.}
\label{ipiprof}
\end{figure*}

\subsubsection{The local excess of the synchrotron emission}

The 408~MHz total intensity map shows a nearly constant high-latitude
($|b|>30\degr$) emission, which can be explained in two ways. 
There might exist an isotropic disk-centered halo emission component, 
which is difficult to model from our location in the Galactic plane far 
from the centre. However, the halo magnetic field we derived from RMs 
makes a smooth total intensity halo emission component quite unlikely.
Otherwise the high-latitude emission can be attributed to local
emission. In fact the observed local synchrotron enhancement requires
a local excess of the CR electron density and/or an increase of the
local random magnetic field component.

Absorption measurements towards distant HII regions located in the
inner Galaxy at 74~MHz yield a synchrotron emissivity of
0.35~K~pc$^{-1}$ to 1.0~K~pc$^{-1}$ for distances larger than about
2~kpc \citep{nhr+06}, which is consistent with the 408~MHz emissivity
from the model of \citet{bkb85} for a typical spectral index $\beta =
-2.5$ for low frequency synchrotron emission \citep{rcls99}. The
22~MHz survey absorption data from HII regions by \citet{rcls99} give
an emissivity of about 30~K~pc$^{-1}$ for distances less than about
1~kpc, which is a factor of about 2 larger than the typical
emissivities obtained by \citet{bkb85}, indicating a local synchrotron
enhancement by a factor of 2. \citet{ft95} summarized low-frequency
absorption measurements up to 2~kpc, which indicate a local
synchrotron enhancement by a factor of about 3. From polarization
modelling towards the very local Taurus molecular cloud \citet{wr04}
confirmed a significant local synchrotron excess.

The solar system is located in a low-density region filled with hot
gas, called the ``local bubble''. Its size is about 200~pc. The
``local bubble'' is believed to be created by about twenty supernovae
from the nearby Scorpius-Centaurus OB clusters during the last 10-15
Myr. The last explosion may have occurred about 1~Myr ago
\citep{bd06,fbd+06}. There is compelling evidence that shock
acceleration in SNRs is the source for energetic CR electrons
\citep[e.g. ][]{kkyn04}. With a typical diffusion coefficient of
10$^{28}$~cm$^2$~s$^{-1}$ \citep{kkyn04} CR electrons may have reached
about 600~pc distance and thus increase the CR electron density in the
local bubble.  Alternatively, multiple fragments with compressed
magnetic fields from the shells of the numerous very old SNRs forming
the ``local bubble'' might have survived in the interstellar medium
and enhance the local synchrotron emission.  From this scenario the
local excess of the synchrotron emission can be achieved either by an
increase of the local CR electron density and/or by an enhanced local
random magnetic field as already stated earlier.  We have tested both
cases and find that the results are the same.  For the simulation
results shown below we assume a local spherical component within 1~kpc
from the Sun, where the CR electron density exceeds the disk component
by a factor 1.5.

\subsubsection{Simulated 408~MHz total intensity and 22.8~GHz 
polarized intensity}

In Figs.~\ref{toti} and \ref{ipiprof} the latitude and longitude
profiles from the simulated maps are compared with those from the
408~MHz and the WMAP 22.8~GHz polarized intensity all-sky surveys.
All three models fit the total intensity profiles quite well. However,
for polarized intensities the BSS magnetic disk field cannot well
reproduce the small longitude asymmetry relative to the Galactic 
centre along the Galactic plane (Fig.~\ref{ipiprof}).

The direction of the halo magnetic field is in the opposite direction
to the disk field above and in the same direction below the Galactic
plane. Therefore the resulting total magnetic field strength is
stronger below than above the Galactic plane. In our model the CR
electron density is truncated at $z$ = 1~kpc and the halo field
decreases quickly towards the Galactic plane, which makes the
resulting intensity difference quite small, e.g. about 20~$\mu$K for
polarized 22.8~GHz emission (see Fig.~\ref{ipiprof}). Therefore this
asymmetry is difficult to prove observationally and requires a careful
absolute zero-level setting for the large-scale emission. For the case
of a smaller halo magnetic field and an extent of the CR electrons
beyond the present limit of 1~kpc in $z$ this difference will remain
small.

\subsubsection{Remarks on the random Galactic magnetic field}

We assume a random field component of uniform strength throughout the
Galaxy.  However, the random magnetic field could be considered to be
stronger in the Galactic plane, where supernova explosions and violent
star formation processes inject energy by shock waves or stellar winds
into the interstellar medium and thus initiate higher fluctuations of
the magnetic field than at high latitudes. From the analysis of the
structure function of EGSs it was shown that fluctuations of the
magnetized interstellar medium in the Galactic plane are
longitude-dependent \citep{sh04} and also different for arm and
inter-arm regions \citep{hgb+06}. Our actual simulations are not
sensitive to spatial variations of the random magnetic field at
high-latitudes because of the present truncation of the CR electrons
at $z$ = 1~kpc.

The local total magnetic field vector $\mathbf{B}_{\rm
tot}=\mathbf{B}+\mathbf{b}$ in our model is about 4~$\mu$G with a
ratio between the random to the regular magnetic field component of
about 1.5. We note that this total field strength is just consistent
with the value of 6$\pm$2~$\mu$G, which was derived by assuming energy
equipartition between CRs and magnetic fields from the synchrotron
model of \citet{bkb85} \citep[Berkhuijsen, personal communication,
shown in Fig.~1 in ][]{b01}.

\subsection{Depolarization}

Using the Galactic magnetic field model and the CR electron
distribution we are able to model the observed total intensity map at
408~MHz and the polarized intensity map at 22.8~GHz. The 1.4~GHz total
intensity map has a different amount of thermal and non-thermal
components and is also well modelled (not shown here). The 1.4~GHz
polarized intensity originates from synchrotron emission modulated by
the interstellar medium and should be reproduced as well after
accounting for depolarization effects, which needs to be discussed in
some detail.

\subsubsection{Simulation problems}

\begin{figure}[!htbp]
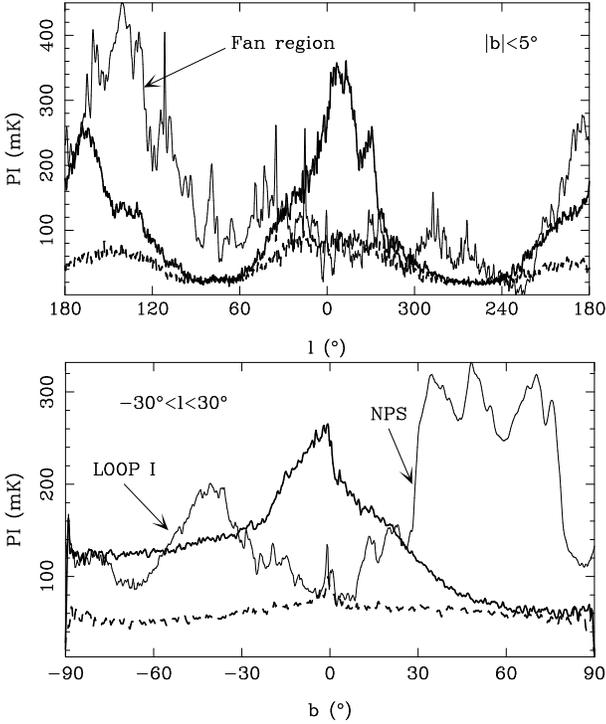

\includegraphics[angle=-90,width=8cm]{pi.l.l.ps}
\includegraphics[angle=-90,width=8cm]{pi.l.b.ps}
\caption{Simulated polarized intensities at 1.4~GHz (shown as thick 
solid and dashed lines) along Galactic longitude in the plane 
(upper panel) and along Galactic latitude (lower panel) are shown 
in comparison to the 1.4~GHz observations (thin solid lines). 
The simulation shown in dashed lines includes a correction for RM 
fluctuations based on the thermal electron filling factor and a related
magnetic field enhancement. For details 
see text. The angular resolution is always $36\arcmin$.}
\label{pi.l}
\end{figure}

We show the simulated results for polarized intensity at 1.4~GHz in
Fig.~\ref{pi.l}, where the longitude profiles along the Galactic plane
and the latitude profiles for $|l|<30\degr$ are presented. The
observed profiles show strong depolarization observed in particular
towards the inner Galaxy (see Fig.~\ref{pik}). Note that prominent
large structures like the "Fan region" and Loop I (as indicated in
Fig.~\ref{pi.l}) are not included in our modelling, as they are
considered as local structures.  Therefore the lower envelopes of the
observed profiles have to be compared with the simulations. Towards
longitudes $l=\pm90\degr$ the magnetic field direction is almost
tangential to the line of sight, which makes any polarized intensity
intrinsically small and the simulated polarized intensities at 1.4~GHz
are almost consistent with the observed ones. However, towards the
centre and the anti-centre region the simulated polarized intensity is
in general much larger than that observed.

\begin{figure}[!htbp]
\includegraphics[angle=-90,width=8cm]{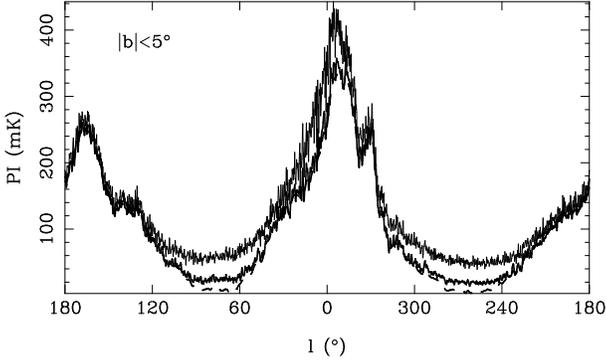}
\caption{Simulated polarized intensities at 1.4~GHz along the Galactic
plane are shown at 15$\arcmin$ (thin solid line) and after smoothing to
36$\arcmin$ (thick solid line) and 2$\degr$ (dashed line).}
\label{sm}
\end{figure}

We show in the following that the random magnetic field component is
properly set in our simulation: Assuming that the intrinsic random
field component is located in cells with a common size $l$, the number
of cells in the $i$-th volume unit calculates then as $N_c^i=i^2\Delta
r^3\Delta\theta^2/l^3$. In the outer (large $i$) unit the number of
cells is very large. However, in the simulation we specify also for
the outer unit just one direction for the random magnetic field
component, which means the number of random magnetic field cells is
clearly underestimated. This way the simulated polarized emission may
be too large.  If this is the case the simulated polarized emission
should significantly decrease when the maps are smoothed. We smoothed
the simulated polarization map to an angular resolution of
36$\arcmin$, which is the resolution of the observed 1.4~GHz all-sky
map, and to a resolution of 2$\degr$. The results are shown in
Fig.~\ref{sm} for a cut along the Galactic plane. It is evident that
the changes in the profile after smoothing are small, which indicates
that the cell size for the random field component is small enough even
for large radii.  This proves that the random field component to be
properly considered in the simulations.

\subsubsection{Solutions to increase the depolarization}

As shown by \citet{sbs+98} Faraday depolarization effects become
strong when either the RM or the RM fluctuations along the line of
sight ($\sigma_{\rm RM}$) increases. For our simulations the RM
fluctuations can be estimated: $\sigma_{\rm
RM}\approx0.81<{n_e}>\sigma_b\sqrt{r\Delta r}$, where $<{n_e}>$ is the
average thermal electron density. In our calculation $\Delta
r=20$~pc. For typical values such as $r=10$~kpc,
$<{n_e}>=0.03$~cm$^{-3}$, $\sigma_b=3$~$\mu$G, the RM fluctuation can
be estimated to be about 32~rad~m$^{-2}$. This fluctuation should 
produce enough depolarization at 1.4 GHz based on the formula by
\citet{sbs+98}.  However, the assumption of identical distributions of
all constituents of the magnetic-ionic medium along the line of sight
made by \citet{sbs+98} clearly differs from the more realistic
properties used for our simulation.  The observed depolarization at
1.4~GHz suggests that this RM fluctuation is too small to cause
significant depolarization.
 
The mean RM has already been constrained by extragalactic sources. We
try to increase the RM fluctuations in each volume unit to introduce a
larger depolarization. We rewrite the formula for RM fluctuations for
the case of a clumpy thermal medium using the same units as in
Eq. (1), which reads:

\begin{equation}
\sigma_{\rm RM}^2=
0.81^2<n_e^2b^2>r\Delta r=0.81^2\frac{<n_e>^2\sigma_b^2}{f_{nb}}r\Delta r, 
\end{equation}
where the factor $f_{nb}$ is defined as 
\begin{equation}
f_{nb}\equiv\frac{<n_e>^2\sigma_b^2}{<n_e^2b^2>}
\end{equation}

\begin{figure}[!htbp]
\includegraphics[angle=-90,width=8cm]{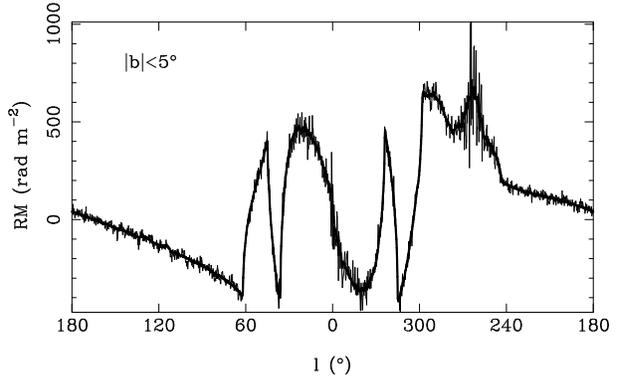}
\caption{The original simulated RM distribution for the ASS+RING disk
field (thin line) and the RM distribution including RM fluctuations by
$f_{nb}=0.01$ (solid line).  }
\label{rm_comp}
\end{figure}

Then we calculate the RMs in each volume unit as
$RM=RM_0+RM_r/\sqrt{f_{nb}}$, where $RM_0$ and $RM_r$ are calculated
from the regular and random magnetic field components, respectively.
From the simulations we found that for $f_{nb}=0.01$ the polarized
intensity in the Galactic plane is significantly reduced and the
simulated results are consistent with the observations, as seen in
Fig.~\ref{pi.l}. The correction does not change the total RM along the
line of sight on average, which we prove in Fig.~\ref{rm_comp},
because just the random RM component in each volume unit is increased.

\begin{figure}[!htbp]
\includegraphics[angle=-90,width=8cm]{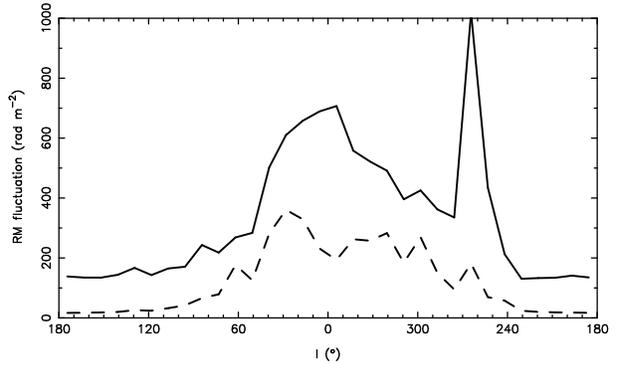}
\caption{The spatial fluctuations of RMs before (dashed) and after
(solid line) considering RM variations ($f_{nb}=0.01$) in a clumpy 
magnetized WIM along the line of sight.}
\label{sigrm}
\end{figure}      

What is the physical meaning of the factor $f_{nb}$ ?  If the thermal
electron density and the random field are independent, i.e.,
$<n_e^2b^2>=<n_e^2><b^2>$, the factor $f_{nb}$ is identical to the
thermal electron filling factor $f_{e}$. Note that $<b^2>=\sigma_b^2$.
The thermal electron filling factor in the plane is about 0.07 from
our simulations, while \citet{bmm06} derived values for $f_{e}$
between 0.04 and 0.07, which is larger than the value $f_{nb}$ = 0.01
as obtained above.  This indicates a coupling between the thermal
electron density and the random magnetic field strength. In general we
may write $f_{nb} = f_ef_c$, where we take $f_c$ as the coupling
factor. In case a dependence like $b^2\propto (n_e^2)^\alpha$ exists,
we have $f_c\propto f_e^\alpha$.  Our simulations clearly favor
$f_{nb}<f_e$ with $\alpha>0$. This means that the thermal electron
density and the random magnetic field strength are
correlated. However, we note that the physics and a theoretical value
for the coupling factor needs to be worked out. Therefore we take our
results as qualitative and did not attempt to account for possible
spatial variations of $f_{nb}$ as it is suggested by $f_{e}(z)$.

\subsubsection{Comment on the observed RM fluctuations}

We finally discuss the observed large fluctuating RMs from EGSs in the
Galactic plane (see Fig.~\ref{rmgp}). Note that these fluctuations are
in principle different from those we discussed above. RM variations of
EGSs result from different ISM properties for each line of sight,
while RM fluctuations causing depolarization are along a single line
of sight. However, larger RM variations are expected for different
line of sights, if the RM fluctuations are large at all. Therefore
both cases are related. As shown in Fig.~\ref{sigrm} the spatial RM
fluctuations are significantly larger after including the factor
$f_{nb}=0.01$. We note that the large fluctuations are now consistent
with the RM observations presented in Fig.~\ref{rmgp}. An important
implication is that the large spatial RM scattering is not due to
fluctuations of large random field components, but to the thermal
filling factor and a coupling factor between the electron density and
the random field strength of the clumpy magnetized WIM.

\subsection{Summary}
\begin{table*}[!htbp]
\caption[]{Observational features versus model constraints based on the 
NE2001 thermal electron model.}
\renewcommand{\arraystretch}{1.15}
\label{model}
\begin{center}
\begin{tabular}{ll}
\hline\hline
observational features &
   model constrains \\
\hline
22.8~GHz free-free emission template and $\beta_{22/408\,\,{\rm MHz}}$ &
    NE2001 plus filling factor for thermal electrons \\
EGS RM profile along the Galactic plane & 
 ASS+RING, ASS+ARM or BSS regular disk field, local field strength 2~$\mu$G\\
High-latitude RMs asymmetry (to the plane and center) & 
    Asymmetric toroidal halo field, field strength up to 10~$\mu$G, CRs up to 1~kpc\\  
RM latitude profile at $100\degr<l<120\degr$ & 
    Favour ASS+RING or ASS+ARM model \\
Longitude PI asymmetry to the centre at 22.8~GHz & 
    Favour ASS+RING or ASS+ARM model \\
Synchrotron emission longitude profiles: 408~MHz I/22.8~GHz PI & 
    Disk CR electron component and 3~$\mu$G random field component\\
Isotropic high-latitude emission at 408~MHz & 
    Local excess of the synchrotron emission \\
Strong depolarization at 1.4~GHz & 
    Large RM fluctuations along the line of sight\\ 
  & by a small $n_e$ filling factor and a coupling of the 
    random field with $n_e$\\
\hline
\end{tabular}
\end{center}
\end{table*}

\begin{figure*}
\begin{minipage}{0.5\textwidth}
\includegraphics[width=8cm]{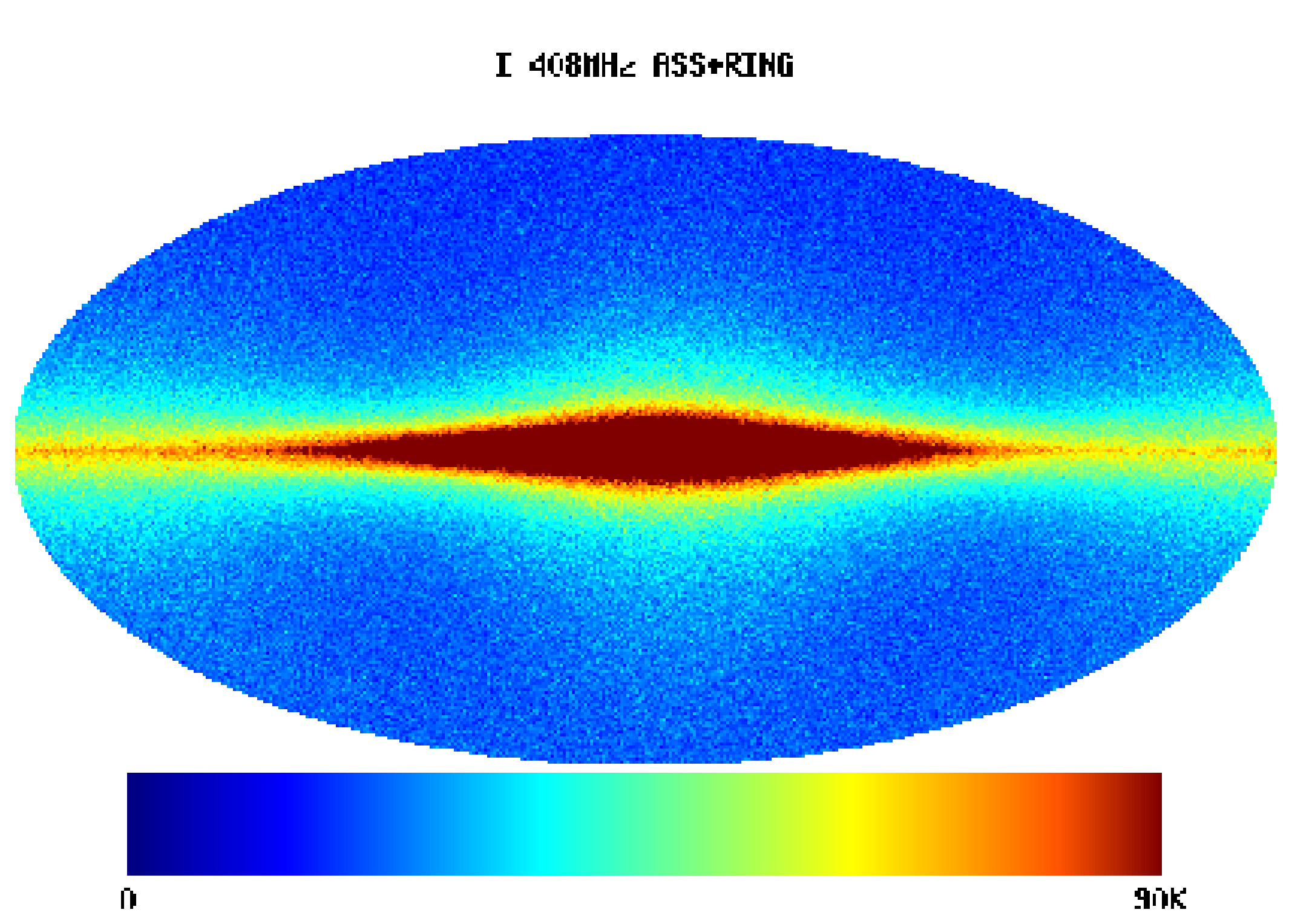}
\end{minipage}
\begin{minipage}{0.5\textwidth}
\includegraphics[width=8cm]{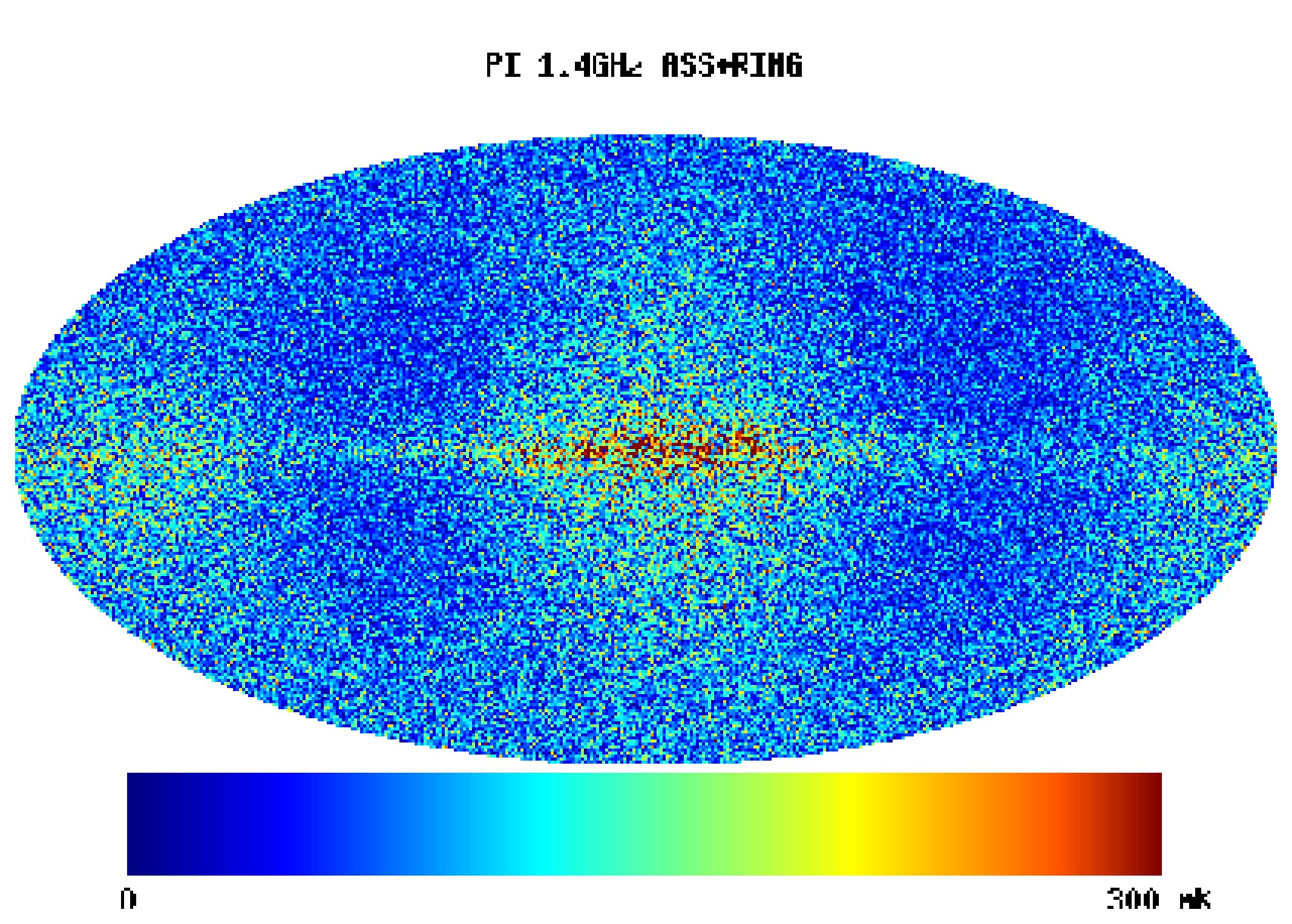}
\end{minipage}
\begin{minipage}{0.5\textwidth}
\includegraphics[width=8cm]{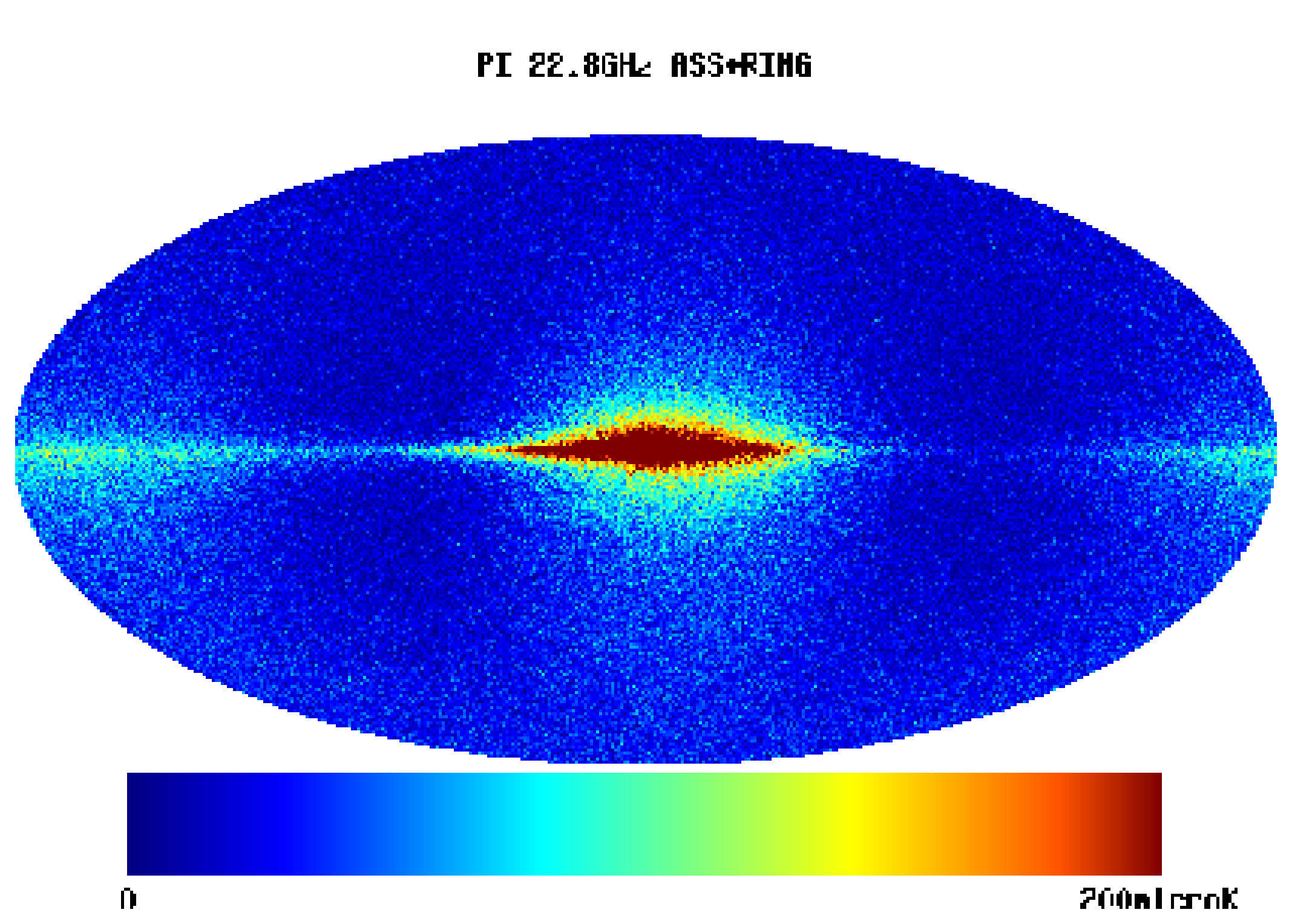}
\end{minipage}
\begin{minipage}{0.5\textwidth}
\includegraphics[width=8cm]{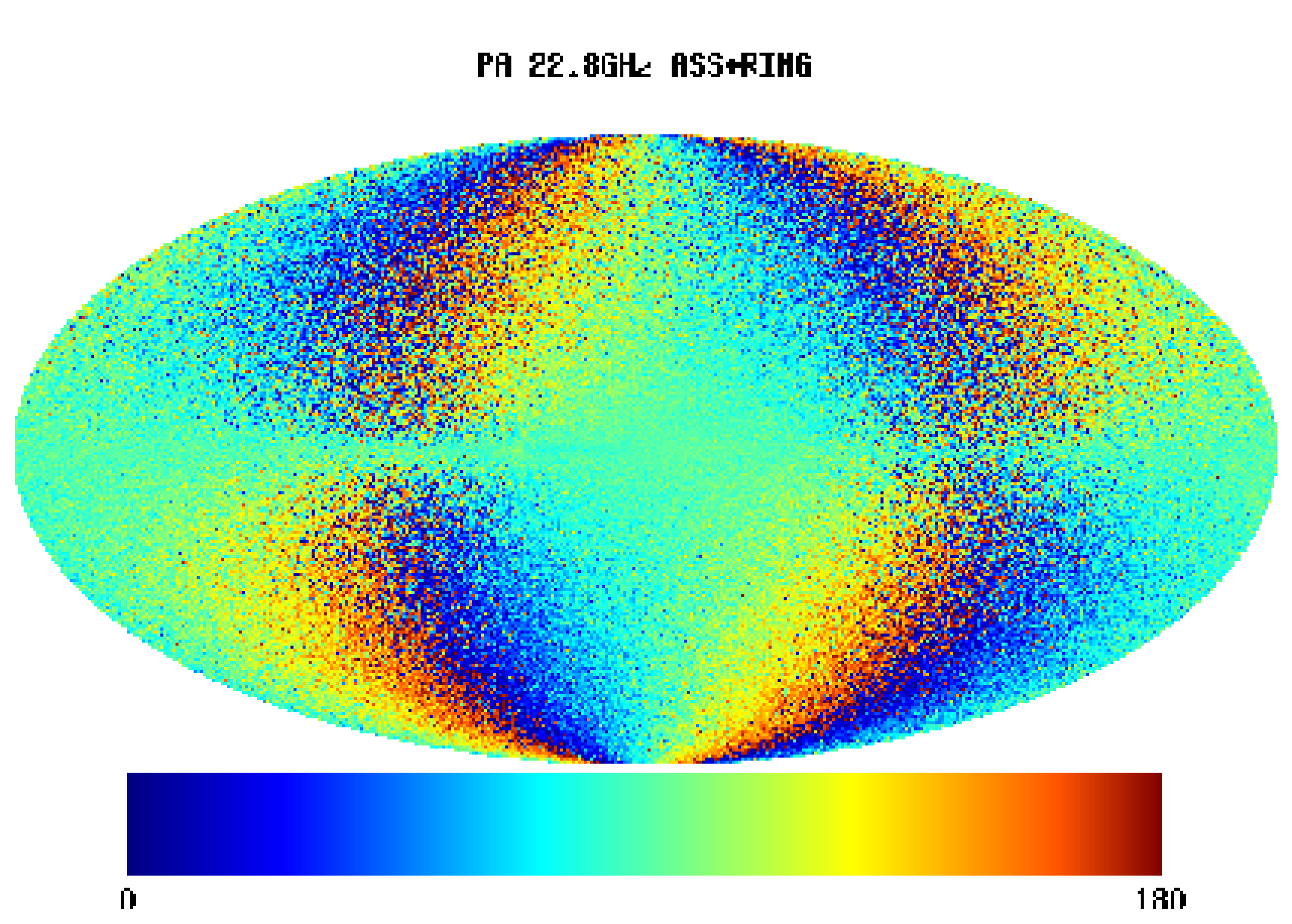}
\end{minipage}
\caption{Simulated all-sky maps for total intensity at 408~MHz,
  polarized intensity at 1.4~GHz and at 22.8~GHz and polarization
  angle at 22.8~GHz are shown for the ASS+RING magnetic field
  configuration. All the maps have a resolution of 15$\arcmin$.}
\label{maps}
\end{figure*}

We summarize the modelling constrains and the related observations in
Table~\ref{model} and show the simulated 408~MHz total intensity, the
1.4~GHz polarized intensity, 22.8~GHz polarized intensity and
polarization angle at 22.8~GHz in Fig.~\ref{maps} based on the
ASS+RING disk field configuration. All maps shown have an angular
resolution of 15$\arcmin$ and can be obtained on request from 
the authors. 

\section{Discussion}

\subsection{Is a large halo field physically reasonable?}

Halo magnetic fields were detected just in a handful of nearby
galaxies such as NGC 891 with a halo field running parallel to the
plane or NGC 4631 showing a vertical field direction.  The field
strength is between 4~$\mu$G and 6.5~$\mu$G for $z$ between 2~kpc to
5~kpc \citep[e.g.][]{dkwk95}.  However, the observed high-latitude RMs
require a stronger halo field for the Galaxy as shown above.

In case the dynamo theory is relevant to describe the Galactic 
magnetic field evolution we learn that in a thick disk a asymmetric 
halo field can be generated as shown by \citet{ss90}. Simulations by
\citet{bdm+92} indicate that it takes a long time, which is 
comparable to the Hubble time, that the halo field reaches stability. 

Considering hydrostatic equilibrium in the halo a mid-plane magnetic
field strength of about 4~$\mu$G with a scale height of 4.4~kpc in the
solar neighbourhood is needed \citep{kk98}. The present halo model
gives a field strength of about 7~$\mu$G at $z$ of 1.5~kpc. This field
sharply decreases towards the Galactic plane. Towards larger $z$ our
model is insensitive to any scale height.

\subsection{The truncation of the CR disk electrons at 1~kpc}

CR electrons gyrate in a magnetic field with a radius
$r_g\approx1\times10^{-6}E/B$~pc, where the energy $E$ is in units of
GeV and the magnetic field strength is in $\mu$G
\citep{fer01}. Turbulence in the halo can be inferred from
super-bubbles generating shock-waves from supernovae in OB clusters
and stellar winds. The outer scale can be estimated to be about
100~pc, which is much larger than the gyrating radius $r_g$. In that
case diffusion of CRs both along and across magnetic field lines is
strong. The diffusion along magnetic field lines is much larger when
the turbulent component is smaller than the regular field component
\citep{clp01} as it is the case for the present halo model. Although
the diffusion of CRs in a turbulent field is still debated
\citep{ptu06}, it seems unlikely that the CR electrons can be
prevented to diffuse out of the Galactic plane into the region with
the large halo field.  Therefore the truncation of the CR electron
density at 1~kpc we need for our simulations seems unrealistic.

However, if we give up the 1~kpc CR electron truncation the modelled
synchrotron emission will increase largely at high
latitudes. As a consequence two ridges of strong emission running
parallel to the Galactic plane will show up, which are not
observed. In case we reduce the halo field to avoid the synchrotron
emission problem, we are unable to model the RM data at high
latitudes. To avoid this it is required to increase the thermal
electron density or the scale height of the DIG.

\subsection{The scale height of the DIG}

The scale height of the DIG of about 1~kpc results from fitting
$DM\sin|b|$ versus $|z|$ for pulsars with independent distance
measurements \citep{gbc01,cl02}. In fact the most stringent
constraints come from distant pulsars located in globular
clusters. However, there are just very few clusters with a large $z$
and the ISM has many voids and clumps, which may influence the scale
height derived from limited data.

From the WHAM H$\alpha$ survey observations in the Perseus arm
direction \citep{hrt99,rht99} the H$\alpha$ intensity can be fitted
with a scale height of 1~kpc and a mid-plane density of
0.06~cm$^{-3}$.  This electron density is larger than the mid-plane
density of about 0.03~cm$^{-3}$ in the NE2001 model. Since the NE2001
model fits the $DM\sin|b|$--$|z|$ relation very well, this might
indicate that the electron density at high $z$ could be larger.  The
WHAM survey clearly shows WIM emission enhancements for large $|b|$
relative to the $EM\sin|b|$ distribution \citep{hrt99,rht99}, which
indicates the presence of areas of large electron densities at high
latitudes.  However, how these high-latitude features are distributed
is rather unclear at present.

In order to check how the high-latitude electron distribution
influences the present simulations we run the HAMMURABI code by
increasing the scale height of the DIG to 2~kpc by keeping the
mid-plane electron density from the NE2001 model unchanged. As a
consequence from this model a regular halo field magnetic strength of
about 2~$\mu$G with a scale height of 4~kpc is required, which sounds
more reasonable compared to the 10~$\mu$G magnetic field strength
enforced by a 1~kpc DIG scale height. In this case the halo field 
parameters agree with the estimates of \citet{ss90} based on the dynamo 
theory. The 1~kpc truncation of CR electron distribution could also be 
avoided that way. We conclude that the high-latitude thermal electron 
density is a critical issue for Galactic 3D-modelling and requires more 
investigations.

\section{Conclusions}

Galactic 3D-models of the distribution of thermal electrons,
cosmic-ray electrons and magnetic fields were used to calculate via
the HAMMURABI code all-sky total intensity, polarization and RM
maps. A comparison with observed maps over a wide frequency range is
made to adapt the 3D-model parameter in the best possible, although
still qualitative way (Table~1).

The main conclusions are :
\begin{itemize}
\item The NE2001 thermal electron model combined with a filling 
factor $f_e$ as derived by \citet{bmm06} reproduces the optical 
thin 22.8~GHz free-free emission template and also the observed spectral 
flattening in the Galactic plane between 408~MHz and 22~MHz caused by
absorption of optically thick thermal gas. We find indications that the 
filling factor might be larger towards high $z$ than that found by 
\citet{bmm06}.

\item An ASS field configuration plus a reversal inside the solar circle 
either in a ring or between the inner edges of the Sagittarius-Carina 
arm and the Scutum-Crux arm is favoured by RMs of EGSs observed along the 
Galactic plane. A BSS field configuration also fits these RMs, however, it 
then fails to fit the observed RM gradient in latitude direction. The 
strength of the regular magnetic field component at the solar radius is 
2~$\mu$G. The field in the Galactic centre region is constant at 2~$\mu$G in 
a direction opposite to that at the solar radius. 

\item The strength of the Galactic random field component is 3~$\mu$G.

\item A regular halo magnetic field with a maximum of 10~$\mu$G at a height 
of 1.5~kpc is required to account for the observed high-latitude RMs. This 
toroidal field is asymmetric to the Galactic plane and decreases sharply 
towards the plane. The variation of the field towards high $z$ is not a 
sensitive parameter in our simulation. 

\item The CR electron distribution has an exponential scale height
 of 1~kpc and a radial scale length of 8~kpc. The disk is truncated at a 
height of 1~kpc. The local CR flux density at 10~GeV is taken twice the 
directly measured value at earth. The energy power law spectral index 
$p$ is taken to be constant at $-$3 for frequencies above 408~MHz and 
$-$2 below. 

\item The observed local excess of synchrotron emission is taken into 
account. The excess is either caused by an enhanced local CR electron 
density or by an increase of the local turbulent magnetic field. Our 
modelling can not distinguish between both possibilities. 

\item In addition to the filling factor we introduce a coupling factor 
between the thermal electron density and the random magnetic field 
component to reproduce the observed large RM fluctuations and the observed 
depolarization at 1.4~GHz.
\end{itemize}

However, a number of caveats remain to the above conclusions.  In our
simulation we use the NE2001 model as input for the thermal electron
distribution and density. This requires a large halo field strength
and hence a truncation of the CR electrons at $z$~=~1~kpc. This might
be not very realistic. Indeed, there are indications as discussed in
Sect. 7.3 that the scale height of thermal electrons might be larger
than 1~kpc. When the thermal electron scale height is increased by a
factor of 2 we get a regular halo magnetic field of 2~$\mu$G and the 
truncation of the CR electron density at 1~kpc is avoided. We conclude 
that the high-latitude thermal electron density needs more investigations 
and thus the halo field parameters remain open.

For large areas along the Galactic plane RM data are still missing or
very rare. A final answer on the magnetic field configuration depends
on a much more complete RM data set. Pulsar RM data are not yet
included in our model, although they are potentially better tracers of
the regular Galactic magnetic field along the line of sight than those
from EGSs as they show no intrinsic RM. Pulsar RMs are clearly
important to detail the disk magnetic field throughout the Galaxy on
smaller scales than in our model and might also reveal deviations from
a general regular pattern being assumed here.  Despite of significant
observational progress over the last years \citep[e.g. ][]{hml+06}
more RM data are clearly needed to be conclusive and pulsar distances
need to be better constrained in addition.

We note that our 3D-model differs from previous models, which were
derived from selected observational data.  These models then fail to
properly reproduce other observations, which, however, trace relevant
information of the interstellar medium, which can not be ignored on
the way towards a physical relevant model.

\begin{acknowledgements}
X.H. Sun acknowledges support by the European Community Framework
Programme 6, Square Kilometre Design Study (SKADS), contract no
011938. We like to thank JinLin Han for providing RM data prior to
publication used to create the synthesized RM map shown in Fig.~3 and
used in Fig.~12. We also thank Jo-Anne Brown for providing her
unpublished RM data from the extended CGPS included in Fig.~12.  We
acknowledge discussions with Elly Berkhuijsen and Rainer Beck.  We
thank Rainer Beck and Patricia Reich for careful reading of the
manuscript and in particular the anonymous referee for valuable comments.

\end{acknowledgements}

\bibliographystyle{aa}
\end{document}